\documentclass[a4paper,11pt]{article}

\usepackage{jheppub,bm,color}
\usepackage{subfigure}
\newcommand{\dis}{\displaystyle}
\newcommand{\rv}{{\bm r}}

\newcommand{\dd}{\text{d}}

\newcommand{\kv}{\bm{k}}
\newcommand{\pv}{\bm{p}}
\newcommand{\Bv}{\bm{B}}
\newcommand{\zev}{\bm{0}}
\renewcommand\d{\partial}
\newcommand{\na}{\bm{\nabla}}

\begin{document}
\title{Does the chiral magnetic effect change the dynamic universality class in QCD?}

\author[a]{Masaru Hongo,}
\author[b]{Noriyuki Sogabe}
\author[b]{and Naoki Yamamoto}

\affiliation[a]{iTHEMS Program, RIKEN, Wako 351-0198, Japan}
\affiliation[b]{Department of Physics, Keio University, Yokohama 223-8522, Japan}

\emailAdd{masaru.hongo@riken.jp}
\emailAdd{nori.sogabe@keio.jp}
\emailAdd{nyama@rk.phys.keio.ac.jp}

\abstract
{In QCD matter under an external magnetic field, the chiral magnetic effect (CME) leads to the collective gapless mode called the chiral magnetic wave (CMW). Since dynamic universality class generally depends on low-energy gapless modes, it is nontrivial whether the CME and the resulting CMW change that of the second-order chiral phase transition in QCD. To address this question, we study the critical dynamics near the chiral phase transition in massless two-flavor QCD under an external magnetic field. By performing the dynamic renormalization-group analysis within the $\epsilon$ expansion, we find that the presence of the CME changes the dynamic universality class to that of model A. We also show that the transport coefficient of the CME is not renormalized by the critical fluctuations of the order parameter.}

\maketitle

\section{Introduction}
\label{sec:introduction} 
 
The Beam Energy Scan program at Relativistic Heavy Ion Collider (RHIC) has two main goals: One is the search for anomaly-induced transport phenomena related to quantum anomalies \cite{Adler,BellJackiw}, such as the chiral magnetic effect (CME) \cite{Kharzeev:2007jp,Fukushima:2008xe,Nielsen:1983rb,Vilenkin:1980fu} and the chiral vortical effect (CVE) \cite{Vilenkin:1979ui,Kharzeev:2007tn,Son:2009tf,Landsteiner:2011cp}. The CME and CVE represent the vector currents in relativistic matter with a chirality imbalance along a magnetic field and vorticity, respectively. 
In QCD matter under an external magnetic field (even at zero density), the CME leads to a sound-like density wave called the chiral magnetic wave (CMW) \cite{Kharzeev:2010gd,Newman:2005hd}. It has been argued that the presence of the CMW in the quark-gluon plasmas (QGP) is reflected in the electric quadrupole observables in heavy-ion collisions \cite{Burnier:2011bf}. An experimental signature consistent with the presence of the CMW may have been observed \cite{Adamczyk:2015eqo}, although an alternative explanation and background for the signature, unrelated to the CMW, have been also proposed \cite{Hatta:2015hca,Hongo:2013cqa}.

The other is the search for the QCD critical point(s) (see, e.g., ref.~\cite{Stephanov:2004wx} for a review), whose possible existence has been argued theoretically at high temperature $T$ between the hadron phase and the QGP phase \cite{Asakawa:1989bq,Barducci:1989wi} and at high baryon chemical potential $\mu_{\rm B}$ between the hadron phase and the color superconducting phase \cite{Kitazawa:2002bc,Hatsuda:2006ps,Zhang:2008wx}. In general, remarkable consequences from the existence of a critical point are the universality of the static and dynamic critical phenomena which does not depend on the microscopic details, but only on the symmetries and low-energy gapless modes. It has been previously found that the dynamic universality class of the second-order chiral phase transition in massless two-flavor QCD at finite $T$ and $\mu_{\rm B} = 0$ is equivalent to that of ${\rm O}(4)$ antiferromagnet \cite{Rajagopal:1992qz}, while that of the high-$T$ critical point at finite $\mu_{\rm B}$ in QCD with finite quark masses is the model H \cite{Fujii:2003bz,Fujii:2004jt,Son:2004iv,Minami:2011un} in the Hohenberg-Halperin classification \cite{Hohenberg:1977ym}.%
\footnote{Recently, it has also been found that, unlike the high-$T$ critical point, the high-$\mu_{\rm B}$ critical point belongs to a new dynamic universality class beyond the conventional Hohenberg and Halperin's classification due to the  interplay between the chiral criticality and the presence of the superfluid phonon \cite{Sogabe:2016ywr}.}
However, since QCD under the magnetic field contains the CMW as an additional collective gapless excitation, it is nontrivial whether such a mode changes the dynamic universality class of the second-order chiral phase transition. One can thus ask the following question: \textit{does the CME (and the resulting CMW) change the dynamic universality class in QCD?} 

In this paper, we address this question, based on the dynamic renormalization-group (RG) analysis within the $\epsilon$ expansion, for a well-posed setup that we will describe in section \ref{sec:setup}.%
\footnote{To simplify our discussion, we here ignore the effects of motion of plasmas, where CVE and the resulting gapless collective mode called the chiral vortical wave (CVW) \cite{Jiang:2015cva} are absent, and we shall focus on the possible modifications due to the CME. The extension of our work to include the coupling with the energy-momentum tensor (and CVE and CVW) is deferred to future work.}
Our main results on the dynamic universality classes are summarized in table \ref{tab:DUC}. (Here $C$ is the transport coefficient of the CME defined in eq.~(\ref{eq:PBn5n}) below.) 
In the presence of the magnetic field ${\bm B} \neq {\bm 0}$, but without the CME, the system belongs to the dynamic universality class of model E. 
On the other hand, when the CME is taken into account, the fixed point corresponding to model E becomes unstable, and the dynamic universality class is changed to that of model A.

\begin{table}[t]
\centering
  \begin{tabular}{|c|c|c|}
  \hline $ \Bv={\bm 0},\, C=0 $ \cite{Rajagopal:1992qz} & $\Bv\neq{\bm 0},\, C=0$ & $\Bv\neq{\bm0},\, C\neq0$ \\
\hline
O(4) antiferromagnet &
model E&
model A\\
\hline
  \end{tabular}
  \caption{Summary of dynamic universality classes (massless two-flavor QCD)}
  \label{tab:DUC}
\end{table}

On the way, we also show that the transport coefficient of the CME is not renormalized by the critical fluctuations of the order parameter. Moreover, we find a new dynamic critical behavior that the speed of the CMW tends to zero when the second-order chiral phase transition is approached. This is similar to the phenomenon known as the critical attenuation of sound waves in the liquid-gas phase transition \cite{Minami:2011un,Onuki:1997} and that of first sound waves in the superfluid transition \cite{Pankert:1986}. 
To the best of our knowledge, our work provides the first study of the interplay between the anomaly-induced transport phenomena and dynamic critical phenomena of QCD.

This paper is organized as follows. 
In section~\ref{sec:setup}, we present the setup of our study: symmetries and hydrodynamic variables under consideration. 
In section~\ref{sec:form}, we construct the non-linear generalized Langevin equations describing the QCD critical dynamics in the presence of the CME. 
In section~\ref{sec:RG}, with the help of the so-called Martin-Siggia-Rose-Janssen-de\,Dominicis (MSRJD) path-integral formalism \cite{Martin:1973zz,Janssen:1976,DeDominicis:1978}, 
we perform the dynamic RG analysis to study the static and dynamic critical behaviors within the $\epsilon$ expansion at the one-loop order. 
In section~\ref{sec:CD}, we conclude with discussions. 

In this paper, we set $e = 1$ for simplicity.

\section{Setup}
\label{sec:setup}
We consider two-flavor QCD with massless up and down quarks at finite temperature $T$ and isospin chemical potential $\mu_{\rm I}$ in an external magnetic field ${\bm B}$. 
The effects of finite isospin chemical potential $\mu_{\rm I}$ and axial isospin chemical potential $\mu_{{\rm I}5}$ are implemented
in the quark sector of massless two-flavor QCD Lagrangian as
 \begin{align}
  \mathcal L _{\rm quark} = i \bar q \gamma^\mu D_\mu q + \mu_{\rm I} \bar q  \gamma^0 \tau^3 q +  \mu_{{\rm I}5} \bar q  \gamma^0 \gamma_5  \tau^3 q,
 \end{align}
where $q$ is the quark field, $D_\mu$ is the covariant derivative including gluon fields and an external magnetic field, and $\tau^a$ ($a=1,2,3$) are the generators of ${\rm SU}(2)$ with the normalization ${\rm Tr}(\tau^a \tau^b)= \delta^{ab}/2$. The chemical potentials $\mu_{\rm I}$ and $\mu_{{\rm I}5}$ are coupled to the isospin density $n_{\rm I} = \bar q \gamma^0 \tau^3 q$ and the axial isospin density $n_{{\rm I}5} = \bar q \gamma^0 \gamma_5 \tau^3 q$, respectively. In the following, we will mostly consider the case with $\mu_{\rm I} \neq 0$ and $\mu_{{\rm I}5} = 0$ (except for section~\ref{sec:symmetry}). Note that the CMW appears even at $\mu_{{\rm I}5} = 0$ due to the fluctuation of $n_{{\rm I}5}$ and $n_{\rm I}$.

Some remarks on our setup are in order here. First, we assume massless quarks such that possible quark-mass corrections of the CME can be ignored. Second, we consider finite $\mu_{\rm I}$ instead of baryon chemical potential $\mu_{\rm B}$, because both $n_{\rm I}$ and $n_{{\rm I}5}$ are conserved in massless QCD, and so $\mu_{\rm I}$ and $\mu_{{\rm I}5}$ are well-defined. On the other hand, the conservation of the axial baryon charge $n_{{\rm B}5}$ is violated by the QCD axial anomaly even in massless QCD, and then the axial baryon chemical potential $\mu_{{\rm B}5}$ is not generically well-defined.%
\footnote{One manifestation of this fact is that the wave equation of the CMW for the baryon charge receives correction due to the QCD axial anomaly. However, such an issue does not occur for the isospin charge.} Therefore, in our setup, whether the CME affects the dynamic critical phenomena in QCD becomes a theoretically well-posed question.  

It should also be remarked that, when ${\bm B} = {\bm 0}$ (where the CME is absent), the dynamic universality class of the second-order chiral phase transition in massless QCD at finite $T$ and $\mu_{\rm B}$ is known to be different from that of QCD with finite quark masses, as we mentioned in section~\ref{sec:introduction}: the former is equivalent to the ${\rm O}(4)$ antiferromagnet \cite{Rajagopal:1992qz}\footnote{More precisely, what is studied in ref.~\cite{Rajagopal:1992qz} is the second-order chiral phase transition at $\mu_{\rm B}=0$ and $\mu_{\rm I}=0$. 
The result at finite $\mu_{\rm I}$ can be obtained by setting $\Bv={\bm 0}$ in our analysis, which we will also argue below.
} while the latter is the model H \cite{Fujii:2003bz,Fujii:2004jt,Son:2004iv,Minami:2011un} in the classification by Hohenberg and Halperin \cite{Hohenberg:1977ym}. 
This difference originates from the absence of the pions at low energy below the pion mass and the presence of the mixing between the chiral condensate and the baryon number density in QCD with finite quark masses at finite $\mu_{\rm B}$ \cite{Son:2004iv}.
In this paper, we will address the question as to whether the CME affects the dynamic universality class of massless QCD (but not that of massive QCD).%
\footnote{To study the possible modification of the dynamic critical phenomena in QCD with finite quark masses due to the CME, one first needs to figure out the quark-mass corrections of the CME and CMW. We defer this problem to future work.} 
As we already stated in section \ref{sec:introduction}, we will ignore the CVE and focus on the CME to simplify the following discussion.

\subsection{Symmetries}
\label{sec:symmetry}
We first summarize the chiral symmetry with $\mu_{\rm I}$ and $\mu_{{\rm I}5}$ in the presence or absence of ${\bm B}$ as shown in table~\ref{tab:sym}. 
In the presence of ${\bm B}$, chiral symmetry $\text{SU}(2)_\text L \times \text{SU}(2)_\text R$ is explicitly broken down to its subgroup $\mathcal{G}\equiv\text{U(1)}^3_{\text{V}}\times\text{U(1)}^3_{\text{A}}$ \cite{Shushpanov:1997sf} (irrespectively of the presence of $\mu_{\rm I}$ and/or $\mu_{{\rm I}5}$). 
This symmetry corresponds to the invariance under the following transformation,
\begin{equation}
  q(x) \rightarrow 
  q'(x) =
  e^{i\alpha_{\text{V}} \tau^3} e^{i\alpha_{\text{A}} \tau^3\gamma^5} q (x),
\end{equation}
where $\alpha_{\text{V}}$ and $\alpha_{\text{A}}$ denote the phase-rotating angles associated with $\text{U(1)}^3_{\text{V}}$ and $\text{U(1)}^3_{\text{A}}$ symmetries, respectively. In the particular case, $\mu_{\rm I}=\mu_{{\rm I}5}\neq0$ with $\Bv={\bm 0}$, where the chemical potential is only coupled to right-handed quarks, chiral symmetry is broken only in the right-handed quark sector as $\text{SU}(2)_\text R \rightarrow \text U(1)^3_\text {R}$. 

\begin{table}[t]
\centering
  \begin{tabular}{|c|c|c|c|}
  \hline
& $ \mu_{\rm I}=\mu_{{\rm I}5}=0 $ & $\mu_{\rm I}=\mu_{{\rm I}5}\neq0 $ & $\mu_{\rm I}\neq 0, \, \mu_{{\rm I}5}=0$ \\
\hline
$\Bv={\bm 0}$&$\text{SU}(2)_\text L \times \text{SU}(2)_\text R$&
$\text{SU}(2)_\text L\times \text U(1)^3_\text {R}$&
$\text{U(1)}^3_{\text{V}}\times\text{U(1)}^3_{\text{A}}$
\\
\hline
$\Bv\neq {\bm 0}$&\multicolumn{3}{|c|}{ $\text{U(1)}^3_{\text{V}}\times\text{U(1)}^3_{\text{A}}$} \\
\hline
  \end{tabular}
  \caption{Chiral symmetry of massless two-flavor QCD with $\mu_{\rm I}$ and $\mu_{{\rm I}5}$
  in the presence/absence of ${\bm B}$.}
  \label{tab:sym}
\end{table}

\subsection{Hydrodynamic variables}
\label{sec:HVs} 
While the static universality class near a critical point or second-order phase transition is characterized only by the symmetry breaking pattern of an order parameter, the dynamic universality class is generally affected by the presence of low-energy gapless excitations in addition to the order parameter.
Therefore, it is necessary to identify appropriate gapless degrees of freedom called hydrodynamic variables in the system.
The typical hydrodynamic variables are the fluctuations of the conserved charge densities, the order parameter associated with the critical phenomena, and the Nambu-Goldstone modes associated with spontaneous breaking of some symmetries. Here we will present all the hydrodynamic variables in our setup. 

The first hydrodynamic variable is the order parameter associated with the chiral phase transition, $\Phi_{ij} \sim\bar{q}_j(1-\gamma_5)q_i$ (with $i,j$ being the flavor indices), which can be generally decomposed as 
\begin{align}
\Phi = \sigma\tau^0 + i \eta\tau^0 + \delta^a \tau^a + i \pi^a \tau^a,
\end{align}
where $\sigma = \bar q \tau^0 q$, $\eta = \bar q i\gamma_5 \tau^0 q$, $\delta^a = \bar q \tau^{a} q$, and $\pi^a = \bar q i \gamma^{5} \tau^{a} q$ with $\tau^0 = 1/2$, and we take a sum over repeated indices.
In the absence of the magnetic field ($\bm{B} = {\bm 0}$), $\sigma$ and $\pi^a$ become nearly massless near the second-order chiral phase transition, while $\eta$ and $\delta^a$ acquire finite masses due to the $\text{U(1)}_\text A$ anomaly or the Kobayashi-Maskawa-'t\,Hooft interaction $c(\det \Phi+\det \Phi^\dagger)$ \cite{Pisarski:1983ms}.%
\footnote{In this paper, we assume that $c$ is finite near and above the chiral phase transition at finite $T$.} 
When the system is put under the magnetic field ($\Bv \neq {\bm 0}$), chiral symmetry is explicitly broken down to its subgroup $\mathcal{G}$ (see section~\ref{sec:symmetry}), and the charged pions $\pi^{1,2}$ acquire a mass proportional to $\sqrt{ |\Bv|}$. Hence, in order to study the critical phenomena at long distance and long time scale much larger than $1/\sqrt{ |\Bv|}$, it is sufficient to only focus on $\sigma$ and $\pi^3$ among these variables. We parametrize them by using the two-component order parameter field $\phi_\alpha\ (\alpha=1,2)$, where we defined $\phi_1\equiv\sigma$ and $\phi_2\equiv \pi^3$. 

The second hydrodynamic variable is the conserved charge densities. In this paper, we only take into account the conserved charge densities associated with the symmetry $\mathcal G$ which are coupled to $\phi_\alpha$, i.e., $n_{\rm I}$ and $n_{{\rm I}5}$. Note that although the energy and momentum densities can also couple to these hydrodynamic variables, we only focus on the above hydrodynamic variables. In other words, we consider the situation where the motion of plasmas is frozen and ignore the possible contributions from the dynamics of the energy-momentum tensor $T^{\mu \nu}$.

In the following, $n_{\rm I}$, $n_{{\rm I}5}$, $\mu_{\rm I}$ and $\mu_{{\rm I}5}$ will be abbreviated as $n$, $n_5$, $\mu$ and $\mu_5$ (by suppressing the index I) for notational simplicity.

\section{Formulation}
\label{sec:form}

\subsection{Generalized Langevin equations}
\label{sec:eom}
Generalized Langevin equations for hydrodynamic variables provide the low-energy effective description of a system with dissipation. This effective theory is based on the derivative expansions controlled by small parameters $p \xi \ll 1$ and $\omega \xi\ll1$ with $p \equiv |\pv|$ and $\omega$ being the strength of a characteristic momentum and frequency, respectively, and $\xi$ being the microscopic correlation length of the order parameter. In this paper, we consider a weak magnetic field regime, and take ${\bm B} = O(p)$.
Following the standard procedure (see, e.g., refs.~\cite{Chaikin,Forster}), one can write down the generalized Langevin equations for the hydrodynamic variables $\phi_\alpha$, $n$, and $n_5$ at finite $T$, $\mu$, and ${\bm B}$ as 
\begin{align}
 \label{eq:Langephi}
 \frac{\partial \phi_\alpha(\rv,t)}{\partial t}
 &= -\Gamma\frac{\delta F}{\delta \phi_\alpha (\bm{r},t)} 
 - g\int \dd \rv' \left[\phi_\alpha(\rv,t),n_5 (\rv',t) \right] 
 \frac{\delta F}{\delta n_5 (\rv',t)} 
 + \xi_\alpha(\rv,t),\\
 \label{eq:Langen}
 \frac{\partial n(\rv,t)}{\partial t}
 &= \lambda \na^2\frac{\delta F}{\delta n (\bm{r},t)}
 -\int \dd \rv' \left[n (\rv,t), n_5(\rv',t) \right] \frac{\delta F}{\delta n_5 (\rv',t)}
 + \zeta(\rv,t),\\
 \label{eq:Langen5}
 \frac{\partial n_5(\rv,t)}{\partial t}
 &= \lambda_5\na^2\frac{\delta F}{\delta n_5 (\bm{r},t)}
  - g\int \dd \rv' \left[n_5 (\rv,t), \phi_\alpha(\rv',t) \right] \frac{\delta F}{\delta \phi_\alpha (\rv',t)} \notag\\
&\hspace{84pt} - \int \dd \rv' \left[n_5 (\rv,t), n(\rv',t) \right] \frac{\delta F}{\delta n (\rv',t)}
 + \zeta_5(\rv,t),
\end{align}
where $[A,B]$ denotes a Poisson bracket describing reversible terms. Here the Ginzburg-Landau-Wilson free energy $F$ is given by%
\footnote{Note that $F$ is a functional of charge densities $n,n_5$ at finite fixed $\mu$. One can also write down the free energy $F'$ which is a functional of $\mu$ and is connected to $F$ by the Legendre transformation: $F=F'-\int \dd \rv \mu \bar n$ with $\bar n\equiv \delta F'/\delta \mu$. While $F'$ can involve linear terms of $\mu$, such as $-\int \dd \rv \mu \varepsilon_{\alpha\beta}\Bv \cdot(\na \phi_\alpha)\phi_\beta$, $F$ does not contain such terms, because of the cancellation due to the Legendre transformation.}
\begin{align}
 \label{eq:GL}
 F&=\int \dd \bm{r} \left[ \frac{r}{2}(\phi_\alpha)^2+
 \frac{1}{2}({\bm \nabla}\phi_\alpha)^{2}+u(\phi_\alpha)^2(\phi_\beta)^2+\frac{1}{2\chi}n^2+\frac{1}{2\chi_5}n_5^2+\gamma n\phi_\alpha^2\right].
\end{align}
Summations over repeated indices are understood. Here, $\Gamma,\lambda$ and $\lambda_5$ are the kinetic coefficients ($\lambda$ and $\lambda_5$ denote the conductivities for the ${\rm U}(1)^{3}_{\rm V,A}$ charges), $g$ is the coupling constant between $\phi_\alpha$ and  $n_5$, and $r,u$ and $\gamma$ are some functions of $T$ and $\mu$. The isospin and axial isospin susceptibilities, $\chi$ and $\chi_5$, are defined as the $a=b=3$ components of the generalized susceptibilities, 
\begin{align}
 \label{chi}
 \chi_{ab} \equiv \frac{\partial n_a}{\partial \mu^b}, \qquad 
 \chi_{5,ab} \equiv \frac{\partial n_{5,a}}{\partial \mu^b_5},
\end{align}
where $n_a = n_{\mathrm{R},a} + n_{\mathrm{L},a},\,
n_{5,a} = n_{\mathrm{R},a} - n_{\mathrm{L},a},\,
\mu^a=(\mu^a_\text R+\mu^a_\text L)/2,\,\mu^a_5=(\mu^a_\text R-\mu^a_\text L)/2$.%
\footnote{One might naively think that, in the QGP with chiral symmetry restoration, right- and left-handed sector are decoupled and that $\chi = \chi_5$. In fact, this is true in the presence of ${\rm SU}(2)_{\rm L}$ or ${\rm SU}(2)_{\rm R}$ chiral symmetry. However, this chiral symmetry is explicitly broken down to its subgroup $\mathcal{G}$ at finite $\mu_{\rm I}$ under the magnetic field ${\bm B}$, and this symmetry is not sufficient to ensure $\chi = \chi_5$; see appendix \ref{sec:chi} for the detail. Hence, we assume $\chi \neq \chi_5$ and treat them as independent quantities below. Indeed, we will see that the RG equations for $\chi$ and $\chi_5$ are different at finite $\mu_{\rm I}$ under the magnetic field ${\bm B}$.}
The non-Gaussian term $\gamma n \phi_\alpha^2$, which is forbidden at $\mu = 0$ by the charge conjugation symmetry, can appear at $\mu \neq 0$. 
On the other hand, the term $n_5 \phi^2_\alpha$ is forbidden by the parity symmetry even for $\mu \neq 0$ (when $\mu_5 = 0$). 
The noise terms $\xi_\alpha,\, \zeta$, and $\zeta_5$ are assumed to satisfy the fluctuation-dissipation relations:
\begin{align}
\label{eq:FDROP}
\langle \xi_\alpha(\rv,t) \xi_\beta(\rv',t') \rangle&=2\Gamma\delta_{\alpha\beta}\delta^d(\rv-\rv')\delta(t-t'),\\
\langle \zeta (\rv,t) \zeta (\rv',t') \rangle&=-2\lambda\na^2\delta^d(\rv-\rv')\delta(t-t'),\\
\label{eq:FDRn5}
\langle \zeta_5 (\rv,t) \zeta_5 (\rv',t') \rangle&=-2\lambda_5\na^2\delta^d(\rv-\rv')\delta(t-t'),
\end{align}
and $\langle \xi_\alpha \zeta \rangle = \langle \xi_\alpha \zeta_5 \rangle= \langle \zeta \zeta_5 \rangle=0$, where $d$ is the spatial dimension.

The reversible terms in generalized Langevin equations are given by the Poisson brackets between hydrodynamic variables. We here use the corresponding commutation relations for the reversible terms,
\begin{align}
 \label{eq:PBn5phi}
 \left [n_5 (\rv,t),\phi_\alpha(\rv',t) \right]
 &= \varepsilon_{\alpha\beta}\phi_\beta\delta(\rv-\rv'),\\
 \label{eq:PBn5n}
 \left [n (\rv,t),n_5(\rv',t) \right]
 &= C \Bv\cdot\na\delta(\rv-\rv').
\end{align}
Here $\varepsilon_{\alpha\beta}$ denotes the anti-symmetric tensor with $\varepsilon_{10}=1$, and $C=1/(2\pi^2)$ denotes the coefficient of the CME, which is related to the anomaly coefficient away from the second-order chiral phase transition.%
\footnote{Note that it is a nontrivial question whether the CME coefficient is exactly fixed by the anomaly coefficient even at the second-order chiral phase transition where $\sigma$ becomes massless. This is because fluctuations of massless $\sigma$ can potentially renormalize the CME coefficient. In section~\ref{sec:dynamics}, we will show that the CME coefficient does not receive renormalization even at the second-order chiral phase transition.} 
Equation~\eqref{eq:PBn5phi} can be understood as the classical limit of the corresponding quantum commutations. 
The anomalous commutation relation \eqref{eq:PBn5n} is related to triangle anomalies in quantum field theories \cite{Jackiw, Faddeev:1984jp} and the CME \cite{Son:2012wh}.

\subsection{Dynamic perturbation theory}
\label{sec:MSR}

\subsubsection{Martin-Siggia-Rose-Janssen-de\,Dominicis (MSRJD) formalism}
In order to apply the RG analysis with perturbative calculations to the classical stochastic dynamics considered in the previous section, we convert the generalized Langevin equation into the path-integral formalism. This formulation has been originally developed by Martin-Siggia-Rose, Janssen, and de\,Dominicis \cite{Martin:1973zz,Janssen:1976,DeDominicis:1978}. 
Here we briefly review their formulation following ref.~\cite{Tauber}. We start from generalized Langevin equations of the form: 
\begin{align}
\label{eq:generalLangevin}
 \frac{\partial \psi_N(\rv{,t})}{\partial t}
 &=\mathcal{F}_N[\{\psi_M\}] + \eta_N(\rv,t) ,\\
\label{eq:generalFDR}
\langle \eta_M(\rv,t)\eta_N(\rv',t')\rangle&=L_{MN}(\na)\delta^d(\rv-\rv')\delta(t-t'),
\end{align}
where eq.~(\ref{eq:generalLangevin}) describes the time evolution of hydrodynamic variables in the presence of the noise variables:
\begin{align}
\psi_{N}\equiv \{\phi_\alpha(\rv,t),n(\rv,t),n_5(\rv,t) \},
\quad\eta_N \equiv \{ \xi_\alpha(\rv,t),\zeta(\rv,t),\zeta_5(\rv,t)\},
\end{align}
in our case. 
Here $\mathcal{F}_N$ denotes all the terms which may involve a set of hydrodynamic variables. The noise variables $\eta_N$ are assumed to satisfy the fluctuation-dissipation relation (\ref{eq:generalFDR}) with $L_{MN}$ being a matrix which also contains  $\na$: 
\begin{align}
L_{MN} = \mathrm{diag} (2\Gamma, 2\Gamma, - 2\lambda \na^2, -2 \lambda_5 \na^2),
\end{align}
for our hydrodynamic variables.
One can see that eqs.~(\ref{eq:Langephi})--(\ref{eq:Langen5}) and (\ref{eq:FDROP})--(\ref{eq:FDRn5}) correspond to eqs.~(\ref{eq:generalLangevin}) and (\ref{eq:generalFDR}), respectively. 

In order to translate the Langevin equation into the corresponding path-integral formalism, we consider following correlation functions of the  hydrodynamic variables under various configurations of the noise variables:
\begin{align}
\label{eq:Opsibar}
\left< \mathcal O[\bar \psi] \right> = 
 \mathcal{N} \int \mathcal D [\eta] \mathcal O[\bar \psi]  \exp \left[ 
 - \frac{1}{4} \int \dd t \int \dd \rv \, \eta_M L_{MN}^{-1} \eta_N 
 \right],
\end{align}
where $\bar\psi_N$ denotes the solution of the Langevin equations, and the noises are assumed to obey the Gaussian white noise with a normalization factor $\mathcal{N}$.
Here, summation over repeated indices $M$, $N$ is implied. Then, we insert the following identity into the right hand side of eq.~(\ref{eq:Opsibar}) in order to carry out the path integral of the noise variables in the presence of the noise-dependent 
variable $\mathcal{O}[\bar\psi]$,
\begin{align}
\label{eq:identMSR}
1= \int \mathcal D [\psi] \prod_{N}\delta(\psi_N-\bar \psi_N) = \int \mathcal D [\psi]\,   \prod_N \prod_{\rv,t}  \delta \left( \frac{\partial \psi_N}{\partial t} - \mathcal F_N[\{\psi_M\}]-\eta_N \right).
\end{align}
We have omitted the Jacobian $\det \left( \partial_t - \delta \mathcal F / \delta \psi \right)$ in the right-hand side. This is justified by getting rid of unnecessary graphs containing the so-called \textit{closed response loops} in diagrammatic calculations (see appendix \ref{sec:Jacobian} for more details). Then, we can replace $\mathcal{O}[\bar \psi]$ by $\mathcal{O}[\psi]$ and integrate out the noise variables. Finally, we get
\begin{align}
\label{eq:Opsi}
\left< \mathcal O[\psi] \right> = \mathcal{N}'\int \mathcal D [i\tilde \psi] \int \mathcal D [\psi]\, \mathcal O [\psi]
\exp \left( - S[\{\tilde\psi_M\},\{\psi_M\}] \right).
\end{align}
Here, we have used the Fourier representation of the delta function and introduced the pure imaginary auxiliary field $\tilde \psi_N$, called the response field, for each hydrodynamic variable $\psi_N $. $\mathcal{N}'$ is also another normalization factor which appears after integrating out the noises, and $S[\{\tilde\psi_M\},\{\psi_M\}]$ is an MSRJD effective action given by
\begin{align}
\label{eq:MSR}
S[\{\tilde\psi_M\},\{\psi_M\}] = \int \dd t \int \dd \rv\, \left[ \tilde\psi_N \left( \frac{\partial \psi_N}{\partial t}  -\mathcal{F}_N[\{\psi_M\}] \right)- \tilde\psi_M L_{MN}(\na)\tilde\psi_N \right].
\end{align}
Then, based on the path-integral technique, we can calculate any correlation functions for the hydrodynamic variables $\psi_N$ and the response fields $\tilde \psi_N$.

Now, following the above procedure for eqs.~(\ref{eq:Langephi})--(\ref{eq:Langen5}) and (\ref{eq:FDROP})--(\ref{eq:FDRn5}), we obtain the path-integral formula with the following MSRJD effective action:
\begin{align}
\label{eq:action}
S= \int \dd t \int \dd \rv \ (\mathcal L _ \phi +\mathcal L _n + \mathcal L _{\phi n}).
\end{align}
The first term represents the kinetic term of the order parameters and their four-point interaction:
\begin{align}
\mathcal L_\phi  &= \tilde{\phi}_{\alpha}\left(\frac{\d }{\d t}+\Gamma(r-\na^{2})\right)\phi_{\alpha}-\Gamma\tilde{\phi}_{\alpha}^2+ 4\Gamma u \tilde{\phi}_\alpha \phi_\alpha \phi_\beta^2\,,
\end{align}
where we have introduced the response fields  $\tilde \phi _ \alpha $ for $\phi_\alpha$. The second term of eq.~(\ref{eq:action}) represents the bilinear part of the conserved charge densities $n$ and $n_5$, which are coupled to each other due to the CME:
\begin{align}
\label{eq:Ln}
\mathcal L _n &=\frac{1}{2}(\tilde n, n, \tilde n_5, n_5)
 \left( \begin{array}{cccc}
2\lambda \na^2   
&  \dis \frac{\d }{\d t}-\frac{\lambda}{\chi} \na^2
&0
&\dis \frac{C}{\chi}_5 \Bv \cdot\na \\
-\dis \frac{\d }{\d t}-\frac{\lambda}{\chi} \na^2
&0
& \dis - \frac{C}{\chi}\Bv\cdot\na
&0 \\
0
& \dis \frac{C}{\chi} \Bv \cdot \na
&2 \lambda_5 \na^2
& \dis \frac{\d }{\d t}-\frac{\lambda_5}{\chi_5} \na^2\\
- \dis \frac{C}{\chi}_5 \Bv \cdot\na  
& 0
& -\dis \frac{\d }{\d t}-\frac{\lambda_5}{\chi_5} \na^2& 0
\end{array}
\right)
\left(
\begin{array}{c}
\tilde n\\
n\\
\tilde n _5\\
n_5
\end {array}
\right),
\end{align}
where $\tilde n$ and $\tilde n _ 5$ are the response fields of $n$ and $n_5$, respectively. 
The third term of eq.~(\ref{eq:action}) represents the three-point interactions between the order parameters and the conserved charge densities:
\begin{align}
\label{eq:LOPCD}
\mathcal L _{\phi n}&=-\frac{g\varepsilon_{\alpha\beta}}{\chi_5}
\left(\tilde{\phi}_\alpha \phi_\beta n_5 +\chi_5 \tilde n_5 (\na^2 \phi_\alpha)\phi_\beta \right)
+2\gamma\Gamma \tilde{\phi}_\alpha \phi_\alpha n
-\gamma\lambda\tilde n (\na^2\phi_\alpha^2)+\gamma C\tilde n _5 \Bv\cdot(\na \phi^2_\alpha).
\end{align}
Here we have two types of interactions: terms proportional to $g$ that originate from the Poisson bracket~(\ref{eq:PBn5phi}) and ones proportional to $\gamma$ that originate from the non-Gaussian term $\gamma n \phi_\alpha^2$ in the Ginzburg-Landau-Wilson free energy (\ref{eq:GL}). The former exists even at $\mu=0$ and gives the couplings between different order parameters; the latter exists as long as $\mu\neq0$ and gives the couplings between the same order parameter components. Note also that the last term of eq.~(\ref{eq:LOPCD}) describes the non-linear interaction, which may potentially generate hydrodynamic loop corrections to the anomaly coefficient $C$.%
\footnote{We will show below that this is not the case by the explicit computation.}

\subsubsection{Feynman rules}
The Feynman rules for the action~(\ref{eq:action}) are in order here. 
The bare propagator of the order parameter, $G ^{0}_{\alpha\beta}$, is obtained by calculating the two-point correlation $\langle \phi_\alpha \tilde \phi_\beta \rangle$ from the Gaussian part of $\mathcal L _\phi$ in momentum space:
\begin{align}
G^{0}_{\alpha\beta} (\bm{k},\omega) = G^{0} (\bm{k},\omega) \delta _{\alpha\beta}&\equiv\frac{\delta _{\alpha\beta}}{-i\omega+\Gamma(r+\bm{k}^2)},
\end{align}
which is diagonal with respect to $\alpha$ and $\beta$. The bare propagator of the conserved fields, $D^{0}_{ij}$, is obtained by calculating two-point correlation $\langle n_i \tilde n_j \rangle$ from $\mathcal  L _n$, where $n_i =\{n,n_5\}$. The inverse matrix of $D^0_{ij}$ has the following expression,
\begin{align}
\label{eq:matrix}
\mathcal [D^0 (\kv,\omega)] ^{-1}=
\left(
\begin{array}{cc}
-i\omega+\dis \frac{\lambda}{\chi}\kv^2
&i \dis \frac{C}{\chi_5} \Bv\cdot\kv
\\
i \dis \frac{C}{\chi}  \Bv\cdot\kv&-i\omega + \dis \frac{\lambda_5}{\chi_5}\kv^2
\end{array}
\right)\,,
\end{align}
which has off-diagonal components with respect to $i$ and $j$ because of the CME. 

\begin{figure}[t]
\centering
\begin{tabular}{cc}
\begin{minipage}{.3\textwidth}
\subfigure[$2\Gamma \delta_{\alpha\beta}$]{ \includegraphics[bb=0 0 88 111,height=3cm]{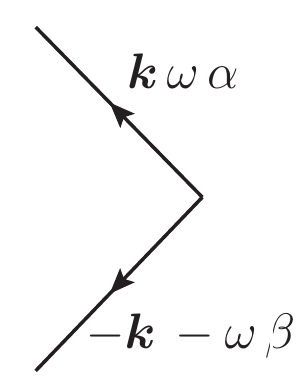}\label{Fig:a}}~
\subfigure[$L^0_{ij}$]{ \includegraphics[bb=0 0 102 112,height=3cm]{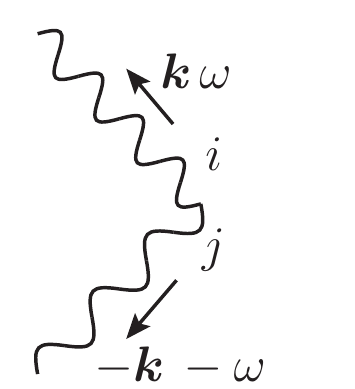}\label{Fig:b}}
\caption{Noise vertices}\label{fig:NV}
\end{minipage}
\end{tabular}\\　\\
\centering
\begin{tabular}{cc}
\begin{minipage}{.7\textwidth}
\subfigure[$U^0_{\alpha ;\beta \gamma \delta}$]{ \includegraphics[bb=0 0 127 112,height=2.8cm]{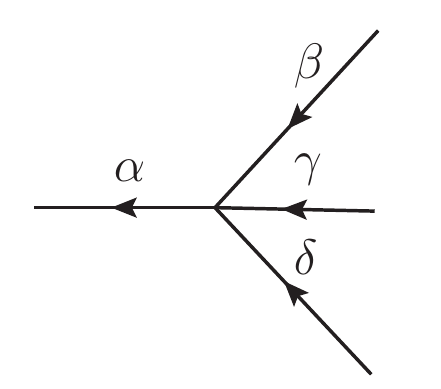}\label{Fig:phi}}~
\subfigure[$V^0_{\alpha;\beta i}$]{ \includegraphics[bb=0 0 115 98,height=2.8cm]{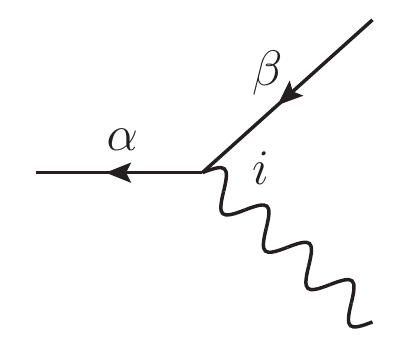}\label{Fig:phin}}~
\subfigure[$V^0_{i;\alpha \beta}$]{ \includegraphics[bb=0 0 140 108,height=2.8cm]{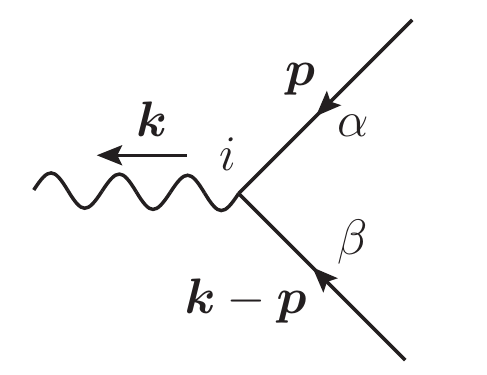}\label{Fig:phin2}}
\caption{Interaction vertices}
\label{fig:IV}
\end{minipage}
\end{tabular}
\end{figure}

The bare noise vertex of the order parameter, $2\Gamma \delta_{\alpha\beta}$, is obtained by calculating two-point correlation $\langle \phi_\alpha\phi_\beta \rangle$ divided by $|G^0|^2$. The bare noise vertex of the conserved charge densities between $n_i$ and $n_j$ is given by 
\begin{align}
L^{0}(\kv)=
\left(
\begin{array}{cc}
2 \lambda \kv^2&0
\\0&2\lambda_5\kv^2
\end{array}
\right), 
\end{align}
which is related to the bare correlation function of the conserved charge densities, $B^0_{ij}$, through the following equation: 
\begin{align}
B^{0}_{ij}(\kv,\omega)&
=D_{il}^{0}(\kv,\omega)L_{lk}^{0}(\kv)[D^{0}(-\kv,-\omega)]^{T}_{kj}=D_{il}^{0}(\kv,\omega)L_{lk}^{0}(\kv)[D^{0}(\kv,\omega)]^{\dagger}_{kj}.
\end{align}
Here, $B^0_{ij}$ is obtained by computing the two-point correlation $\langle n_i n_j \rangle$ from $\mathcal L _n$ as follows:
\begin{align}
B^{0}_{11}(\kv,\omega)&=\frac{ 2\lambda \kv^2 (\omega^2+\lambda_5^2\kv^4/\chi_5^2)+2\lambda_5\kv^2(C\Bv\cdot\kv/\chi_5)^2}{\left|\det[ D^0(\kv,\omega)]^{-1}\right|^2},\\
B^{0}_{12}(\kv,\omega)&=B^{0}_{21}(\kv,\omega)=\frac{2\left(\lambda/\chi+\lambda_5/\chi_5\right) \kv^2 C(\Bv\cdot\kv)\omega}{\left|\det[ D^0(\kv,\omega)]^{-1}\right|^2},\\
B^{0}_{22}(\kv,\omega)&=\frac{2\lambda_5\kv^2(\omega^2+\lambda^2 \kv^4/\chi^2)+2\lambda \kv^2(C \Bv\cdot\kv/\chi)^2 }{\left|\det[ D^0(\kv,\omega)]^{-1}\right|^2}.
\end{align}
The four-point interaction vertex is obtained from $\mathcal L _{\phi}$ as 
\begin{align}
U^0_{\alpha ; \beta \gamma \delta}=-4u \Gamma \delta_{\alpha\beta} \delta_{\gamma \delta},
\end{align}
where the indices $\alpha ;\beta \gamma \delta$ are the shorthand notation of the fields $\tilde \phi _\alpha \phi _\beta \phi _\gamma \phi _\delta$. 
Here and below we write the index $\alpha$ for the response field on the left and the components $\beta, \gamma, \delta$ for the hydrodynamic fields on the right. 
The three-point interaction vertices are obtained as
\begin{align}
\label{eq:bareV}
V^0_{ \alpha; \beta i}
=\left(
\begin{array}{c}
-2\gamma\Gamma\delta_{\alpha \beta}\\
g\varepsilon_{\alpha\beta}/\chi_5
\end{array}
\right),\quad
V^0_{i; \alpha \beta }(\kv,\pv)
=\left(
\begin{array}{c}
-2\gamma\lambda \kv^2 \delta_{\alpha \beta}\\
g[(\kv-\pv)^2-\pv^2]\varepsilon_{\alpha\beta}-2i\gamma C \Bv \cdot \kv \delta_{\alpha \beta}
\end{array}
\right),
\end{align}
where we explicitly write each $i$ component as a vector. 
In the same way as the above, the indices $\alpha; \beta i$ and $i; \alpha \beta $ are the shorthand notation of $\tilde  \phi _\alpha \phi_\beta n_i$ and  $\tilde n_i  \phi _\alpha \phi_\beta $, respectively. The vertex $V^0_{i; \alpha \beta }(\kv,\pv)$ is the function of the outgoing momentum $\kv$ of $n_i$ and the ingoing momentum $\pv$ of $\phi_\alpha$ (see also figure~\ref{Fig:phin2}). 

Diagrammatically, we depict $G^0_{\alpha\beta}$ by the plane line, and $D^{0}_{ij}$ by the wavy line with the outgoing and ingoing components $i$ and $j$. We omit both of the outgoing and ingoing indices $\alpha$ and $\beta$ in the diagram for $G^0_{\alpha\beta}$, because $G^0_{\alpha\beta}$ is diagonal with respect to $\alpha,\beta$. Rather, we shall write $\alpha$ alone at the center of plane lines. Each noise vertex of the order parameters and the conserved charge densities can be understood as the one with two outgoing lines as represented in figure~\ref{fig:NV}.

As is shown in figure~\ref{fig:IV}, each interaction vertex has one outgoing and three or two ingoing lines. Ordinary Feynman rules are applied in figure~\ref{fig:IV} to obtain $n$-point full correlations. Among others, the full propagators $G_{\alpha\beta}$ and $D_{ij}$ are obtained using the self-energies of the order parameter, $\Sigma_{\alpha\beta}$, and those of the conserved charge densities, $\Pi_{ij}$, as
\begin{align}
\label{eq:DysonG}
G^{-1}_{\alpha\beta}(\bm{k},\omega)&=[G^0_{\alpha\beta}(\bm{k},\omega)]^{-1}-\Sigma_{\alpha\beta}(\bm{k},\omega),\\
\label{eq:DysonD}
D^{-1}_{ij}(\bm{k},\omega)&= [D^0_{ij}(\bm{k},\omega)]^{-1} -\Pi_{ij}(\bm{k},\omega).
\end{align}
The three-point vertex function $V_{\alpha;\beta i}(\kv_1,\kv_2,\omega_1,\omega_2)$ can be obtained by computing one-particle irreducible diagrams with outgoing $\tilde \phi _ \alpha$ and ingoing $\phi_\beta,\, n_i$. Here, $\kv_1$ and $\omega_1$ denote the ingoing momentum and frequency of $\phi_\beta$, and $\kv_2$ and $\omega_2$ denote the ingoing momentum and frequency of $n_i$. From the energy and momentum conservation laws, $\tilde \phi_\beta$ has the outgoing momentum $\kv_3 = \kv_1+\kv_2$ and the frequency $\omega_3 = \omega_1+\omega_2$. For later purpose, it is convenient to divide $V_{\alpha;\beta i}$ into its bare contribution $V^0_{\alpha;\beta i}$ and the correction term $ \mathcal V _{\alpha;\beta i}$ as follows:
\begin{align}
\label{eq:mathcalVdef}
V_{\alpha;\beta i}(\kv_1,\kv_2,\omega_1,\omega_2)=V^0_{\alpha;\beta i}+\mathcal V _{\alpha;\beta i}(\kv_1,\kv_2,\omega_1,\omega_2).
\end{align}

\section{Renormalization-group analysis}
\label{sec:RG}

\subsection{Statics}
\label{sec:statics}
We first discuss the static critical behavior by using the RG analysis with $\epsilon$ expansion. The static RG transformation consists of two steps: integrating out the degrees of freedom in the momentum shell between $\Lambda/b$ and $\Lambda$ (with $\Lambda$ being the ultraviolet cutoff and $b$ being some constant larger than unity) and the rescaling:
\begin{align}
\label{eq:scalingcord}
\rv&\rightarrow \rv'=b^{-1}\rv,\\
\label{eq:scalingphi}
\phi_\alpha(\bm{r})&\rightarrow \phi'_\alpha(\bm{r}')=b^a \phi_\alpha(\bm{r}),\\
\tilde{\phi}_\alpha(\bm{r})&\rightarrow \tilde{\phi}'_\alpha(\bm{r}')=b^{\tilde{a}} \tilde{\phi}_\alpha(\bm{r}),\\
n(\bm{r})&\rightarrow n'(\bm{r}')=b^c n(\bm{r}),\\
\tilde{n}(\bm{r})&\rightarrow \tilde{n}'(\bm{r}')=b^{\tilde{c}} \tilde{n}(\bm{r}),\\
n_5(\bm{r})&\rightarrow n'_5(\bm{r}')=b^{c_5} n_5 (\bm{r}),\\
\tilde{n}_5 (\bm{r})&\rightarrow \tilde{n}'_5(\bm{r}')=b^{\tilde{c}_5} \tilde{n}_5(\bm{r}).
\end{align} 
Here $a$, $\tilde a$, $c$, $\tilde c$, $c_5$, and $\tilde c_5$ are some constants, among which $a$, $c$, and $c_5$ will be computed below. 
Hereafter, we work with the spatial dimension $d\equiv4-\epsilon$ with small $\epsilon$, and perform the calculations to leading orders in the expansion of $\epsilon$. 

At the $l$th stage of the renormalization procedure, the static parameters $r_l,\, u_l,\, \chi_l$ and $\gamma_l$ satisfy the same recursion relation as those in model C, i.e., eqs.~(4.5)--(4.8) of ref.~\cite{Halperin:1974zz}.%
\footnote{The susceptibility $\chi$ is denoted as $C$ in ref.~\cite{Halperin:1974zz}.} 
Since there is no non-Gaussian terms for $n_5$ in eq.~(\ref{eq:GL}), $(\chi_5)_l$ is affected only by the trivial scale transformation, and so $(\chi_5)_l$ is finite at the fixed point. 
The recursion relations for the static parameters in the leading-order of $\epsilon$ are 
\begin{align}
r_{l+1}&=b^{d-2a}\{ r_l + 8\bar u _l [\Lambda^2(1-b^{-2})-2r_l \ln b]\}, \\
\label{eq:RRu}
\bar u_{l+1}&=b^{d-4a} \bar u_l \left( 1-40\bar u _l \ln b \right), \\
\label{eq:RRchi}
\chi_{l+1}^{-1}&=b^{d-2c}\chi_l^{-1}(1-4v_l \ln b),\\
\label{eq:RRchi5}
(\chi_5)_{l+1}^{-1}&=b^{d-2c_5}(\chi_5)_l^{-1},\\
\label{eq:RRgamma}
\gamma_{l+1}&=b^{2-c}\gamma_l \left[1- \left( 16 \bar u _l +4v_l \right) \ln b \right].
\end{align}
Here, we introduced the following quantities,
\begin{align}
\label{eq:Parastas}
v\equiv\frac{\gamma^2 \chi \Lambda^{-\epsilon}}{8\pi^2},\quad \bar u\equiv  \frac{u \Lambda^{-\epsilon}}{8\pi^2}\,,
\end{align}
where the factor $1/(8\pi^2)$ is the phase-space volume element in $d = 4$ dimensions.%
\footnote{This factor is denoted as $B$ in ref.~\cite{Halperin:1974zz} and $K_4$ in ref.~\cite{Halperin:1976zza}.}

Let us now compute $a$, $c_5$, and $c$, among which $a$ and $c_5$ can be found immediately. To the order of $\epsilon$, the exponent $a$ is solely determined by the condition that the kinetic term of $\phi_\alpha$ in eq.~(\ref{eq:GL}) is scale invariant:
\begin{align}
\label{eq:a}
a=\frac{d-2}{2}\,.
\end{align}
Recalling the condition that $(\chi_5)_l$ remains finite at the fixed point, we also find from eq.~(\ref{eq:RRchi5}) that
\begin{align}
\label{eq:c5}
c_5=\frac{d}{2}\,.
\end{align}
We then turn to the computation of $c$. Combining eqs.~(\ref{eq:RRchi}) and (\ref{eq:RRgamma}), we obtain the recursion relation for $v_l$,
\begin{align}
\label{eq:RRv}
v_{l+1}&=b^\epsilon v_l \left[1- \left( 32 \bar u_l  + 4v_l  \right) \ln b \right].
\end{align}
The RG equations corresponding to eqs.~(\ref{eq:RRu}) and (\ref{eq:RRv}) become%
\footnote{In general, for the following form of the recursion relation for parameters $A_l$ and $B_l$,
\begin{align}
A_{l+1} = b^{c_A} A_l (1+ B_l \ln b), \nonumber
\end{align}
with some constant $c_A$, one can derive the RG equation for $A_l$ by setting $b = e^l$ and taking the limit $l \rightarrow 0$ as 
\begin{align}
\frac{{\rm d}A_l}{{\rm d}l} = (c_A + B_l) A_l. \nonumber
\end{align}}
\begin{align}
\label{eq:REu}
\frac{{\rm d}\bar u_l}{{\rm d}l} = (d - 4 a - 40 \bar u_l) \bar u_l. 
\\
\label{eq:REv}
\frac{{\rm d}v_l}{{\rm d}l} = (\epsilon - 32 \bar u_l  - 4v_l) v_l.
\end{align}
Then, eqs.~\eqref{eq:REu}-\eqref{eq:REv} tell us the fixed-point values of $\bar u_l$ and $v_l$ as \cite{Halperin:1974zz}%
\footnote{Following the notation of ref.~\cite{Halperin:1976zza}, we use $\bar u_{\infty}$, $v_{\infty}$, etc., for the fixed-point values of $\bar u_l$, $v_l$, etc., rather than $\bar u_*$, $v_*$ as in ref.~\cite{Halperin:1974zz}.} 
\begin{align}
\bar u_\infty= \frac{\epsilon}{40}, \quad
\label{eq:FVv}
\dis v_\infty = \frac{\epsilon}{20}\,.
\end{align}
By returning back to eq.~(\ref{eq:RRchi}), we arrive at
\begin{align}
\label{eq:c}
\dis c=\frac{3d-2}{5}\,.
\end{align}

The static physical parameters near the transition temperature $T_c$ are solely given by the loop-correction terms proportional to $\ln b$ in the static recursion relations \cite{Onuki:PTD}. In particular, the renormalized charge susceptibility $\chi(T)$ is obtained by using eq.~(\ref{eq:RRchi}),
\begin{align}
\label{eq:chiT}
\chi(T)=\chi_0 [1+4v_\infty \ln (\Lambda\xi)]\sim \xi^{\epsilon/5},
\end{align}
where $\chi_0$ is the susceptibility at the cutoff scale $\Lambda$, 
and we used the relation $1+x \ln \Lambda \xi +O(x^2)=(\Lambda \xi)^x$ for $x \ll 1$, by regarding $x=4 v_\infty$ as a small parameter when $\epsilon\ll 1$. 
Defining the critical exponents $\nu$ and $\alpha$ in the standard manner,
\begin{align}
\label{eq:eps}
\xi\sim\tau^{-\nu},\quad\chi\sim\tau^{-\alpha}, 
\end{align}
with $\dis\tau\equiv{(T-T_c)}/{T_c}$ being the reduced temperature, we obtain 
\begin{equation}
\label{eq:alpha/nu}
\dis \frac{\alpha}{\nu}=\frac{\epsilon}{5}\,. 
\end{equation}

The recursion relations at $\mu=0$ can easily be obtained by setting $v=\gamma=0$ in eqs.~(\ref{eq:RRchi})--(\ref{eq:RRgamma}). In this case, one finds $\dis c=c_5={d}/{2}$, so that critical behaviors do not appear in $\chi$ and $\chi_5$.

\subsection{Dynamics}
\label{sec:dynamics}
\subsubsection{Dynamic RG equations and fixed point solutions}
\label{sec:Dynamic RG equations and fixed point solutions}
We now discuss the dynamic critical behavior. 
Similarly to the static RG procedure presented in the previous section, the dynamic RG transformation also consists of two steps: integrating out the intermediate momentum shell, as well as the intermediate frequency $\omega$ from $-\infty$ to $\infty$ and its rescaling: \begin{align}
\omega\rightarrow \omega'=b^{z}\omega, 
\end{align}
where $z$ is the so-called dynamic critical exponent. 

In order to obtain the recursion relations of the dynamic parameters $\Gamma_l,\, \lambda_l,\, \lambda_5,$ and $C_l$ in the MSRJD effective action, we compute the full inverse propagators at the ($l+1$)th renormalization step. For instance, we calculate the full inverse propagator for the order parameter as
\begin{align}
[G_{\alpha\alpha}(\kv',\omega')]^{-1}_{l+1}
&=[G_{\alpha\alpha}(\kv,\omega)]^{-1}_l b^{-\tilde a -a}\notag\\
&=\left [-i\omega \left(1-i\left.   \frac{\d \Sigma_{\alpha\alpha}(\bm{0},\omega)}{\d \omega}   \right|_{\omega\rightarrow0}\right)+\Gamma_l r_l-\Sigma_{\alpha\alpha}(\zev,0) \right.\notag\\
&\qquad\qquad\qquad\qquad\left.
+\left(  \Gamma_l
-\frac{1}{2}\left.   \frac{\d^2 \Sigma_{\alpha\alpha}(\bm{k},0)}{\d \bm{k}^2} \right|_{\bm{k}\rightarrow \bm{0}} \right)   \kv^2 +\cdots
\right]b^{-\tilde a -a}.
\end{align}
Here we used eq.~(\ref{eq:DysonG}) and expanded the diagonal component of the self energy $\Sigma$ with respect to the frequency $\omega$ and wave number $\kv$.
By regarding the term proportional to $\kv^2$ on the right-hand side as $\Gamma_{l+1}\kv'^2$ and including the overall factor $b^{z+d}$ that originates from rescaling the measure of the action, we obtain the recursion relation for $\Gamma_l$,
\begin{align}
\label{eq:RRGamma}
\Gamma_{l+1}&=\Gamma_{l}
\left(
1-
\frac{1}{2\Gamma_l} \left.   \frac{\d^2 \Sigma_{\alpha\alpha}(\bm{k},0)}{\d \bm{k}^2}    \right|_{\bm{k}\rightarrow \bm{0}}
\right)b^{d+z-\tilde a -a-2}.
\end{align}
Furthermore, we regard the term proportional to $-i\omega$ as $-i\omega'$ and include the overall factor, 
which leads to
\begin{align}
\label{eq:atildea}
1&=\left(  1-i \left.   \frac{\d \Sigma_{\alpha\alpha}(\bm{0},\omega)}{\d \omega}   \right|_{\omega\rightarrow0}  \right)b^{d-\tilde a -a}.
\end{align}

In the similar way, we can derive the following recursion relations by computing the inverse propagator for conserved charge densities, $[D_{ij}(\kv',\omega')]^{-1}_{l+1}$, and three-point vertex $[V_{\alpha;\beta i} (\kv_1,\kv_2,\omega_1,\omega_2)]_{l+1}$:
\begin{align}
\label{eq:RRlambdachi}
\frac{\lambda_{l+1}}{\chi_{l+1}}&=\frac{\lambda_l}{\chi_{l}}\left(1-\frac{\chi_l}{2\lambda_l}  \left.    \frac{\d^2 \Pi_{11} (\bm{k},0)}{\d \bm{k}^2}    \right|_{\bm{k}\rightarrow \bm{0}}  \right)b^{d+z-\tilde{c}-c-2},\\
\label{eq:RRlambda5chi5}
\frac{(\lambda_5)_{l+1}}{(\chi_5)_{l+1}}&=\frac{(\lambda_5)_l}{(\chi_5)_l}\left(1-\frac{(\chi_5)_l}{2(\lambda_5)_l}  \left.    \frac{\d^2 \Pi_{22} (\bm{k},0)}{\d \bm{k}^2}    \right|_{\bm{k}\rightarrow \bm{0}}  \right)b^{d+z-\tilde{c}_5-c_5-2},\\
\label{eq:RRCchi}
\frac{C_{l+1}}{\chi_{l+1}}\Bv &=\left(\frac{C_l}{\chi_l}\Bv +\left.   i\frac{\d \Pi_{21}(\bm{k},0)}{\d \bm{k}}    \right|_{\bm{k}\rightarrow \bm{0}}\right)b^{d+z-\tilde{c}_5-c-1},\\
\label{eq:RRCchi5}
\frac{C_{l+1}}{(\chi_5)_{l+1}}\Bv &=\left(\frac{C_l}{(\chi_5)_l}\Bv +\left.   i\frac{\d \Pi_{12}(\bm{k},0)}{\d \bm{k}}    \right|_{\bm{k}\rightarrow \bm{0}}\right)b^{d+z-\tilde{c}-c_5-1},\\
\label{eq:RRgchi5}
\frac{g_{l+1}}{(\chi_5)_{l+1}}\varepsilon_{\alpha\beta}&=\frac{g_l}{(\chi_5)_l} \left(  \varepsilon_{\alpha\beta}  + \frac{(\chi_5)_l}{g_l} 
\left.  \mathcal  V_{\alpha;\beta 2} (\kv_1,\kv_2,\omega_1,\omega_2)     \right|_{\kv_{1,2}\rightarrow \zev, \ \omega_{1,2}\rightarrow 0}\right)b^{d+z-\tilde a -a -c_5},
\end{align}
with the following constraints,
\begin{align}
\label{eq:ctildec}
1&=\left(  1-i \left.   \frac{\d \Pi_{11} (\bm{0},\omega)}{\d \omega}   \right|_{\omega\rightarrow0}  \right)b^{d-\tilde c - c}\,,\\
\label{eq:ctilde5c5}
1&=\left(  1-i \left.   \frac{\d \Pi_{22} (\bm{0},\omega)}{\d \omega}   \right|_{\omega\rightarrow0}  \right)b^{d-\tilde c _5 - c _5}\,.
\end{align}
Therefore, once we can evaluate the self-energy and vertex function corrections, we obtain the recursion relations for the dynamic parameters in the MSRJD action, which results in the dynamic RG equations.

We then perturbatively compute the self-energies, $\Sigma_{\alpha\alpha}(\kv,\omega)$ and $\Pi_{ij}(\kv,\omega)$, and the vertex function $V_{\alpha;\beta2}(\kv_1,\kv_2,\omega_1,\omega_2)$ via diagrammatic expansions. For later convenience, we define the following quantities:
\begin{gather}
\label{eq:Paras}
f \equiv\frac{g ^2 \Lambda^{-\epsilon}}{8\pi^2 \lambda_5 \Gamma},\quad w\equiv\frac{\Gamma \chi}{\lambda},\quad w_5 \equiv\frac{\Gamma \chi_5 }{\lambda_5},\quad h\equiv\frac{C B}{\sqrt{\lambda\lambda_5}\Lambda},\\
\label{eq:Paras2}
X\equiv\frac{2}{\sqrt{(1+w)(1+w_5)+h^2}+\sqrt{(1+w)(1+w_5)}}\sqrt{\frac{1+w}{1+w_5}},\quad X'\equiv \frac{1+w_5}{1+w}X,
\end{gather}
where we introduced $B \equiv |{\bm B}|$. 
According to the detailed analysis in appendix~\ref{sec:calculation}, the recursion relations for the dynamic parameters $\Gamma_l,\, \lambda_l,\, \lambda_5,\, C_l$ and $g_l$ at the one-loop level read
\begin{align}
\label{eq:RRGamma2}
\Gamma_{l+1}&=b^{z-2}\Gamma_{l}
\left[1- (4v_l w_l X'_l-f_l X_l) \ln b \right],\\
\label{eq:RRlambda2}
\lambda_{l+1}&=b^{z-d+2c-2}\lambda_l,\\
\label{eq:RRlambda52}
(\lambda_5)_{l+1}&=b^{z-d+2c_5-2}(\lambda_5)_l \left(  1+\frac{f_l}{2} \ln b \right),\\
\label{eq:RRC}
C_{l+1} &=b^{z+c+c_5-d-1}C_l, \\
\label{eq:RRg}
g_{l+1}&=b^{z - d + c _5}g_l.
\end{align}
Among others, eq.~(\ref{eq:RRC}) shows that the CME coefficient $C$ is not renormalized by the critical fluctuations of the order parameter in this order.
This may be viewed as an extension of the non-renormalization theorem for the CME coefficient at the second-order chiral phase transition where $\sigma$ becomes massless.

The recursion relations for the dynamic parameters (\ref{eq:RRGamma2})--(\ref{eq:RRg}), together with the ones for static parameters (\ref{eq:RRchi})--(\ref{eq:RRgamma}), enable us to obtain the recursion relations for parameters defined in eq.~\eqref{eq:Paras} as
\begin{align}
\label{eq:RRf}
f_{l+1}&=b^{\epsilon}f_l \left[1+ \left(4v_l w_l X'_l-f_l X_l -\frac{1}{2}f_l \right) \ln b \right],\\
\label{eq:RRw5}
(w_5)_{l+1}&=(w_5)_l \left[1 - \left(4v_l w_l X'_l-f_l X_l + \frac{1}{2}f_l \right) \ln b \right],\\
\label{eq:RRw}
w_{l+1}&=w_l \left[1 - \left(4v_l w_l X'_l-f_l X_l - 4 v_l \right) \ln b \right],\\
\label{eq:RRh}
h_{l+1}&=b h_l \left(1- \frac{f_l}{4} \ln b \right).
\end{align}
Then, the dynamic RG equations corresponding to eqs.~(\ref{eq:RRf})--(\ref{eq:RRh}) can be derived in a way similar to 
eqs.~\eqref{eq:REu}-\eqref{eq:REv} as 
\begin{align}
\label{eq:REf}
\frac{{\rm d}f_l}{{\rm d}l}&= \left(\epsilon+ 4v_l w_l X'_l-f_l X_l -\frac{1}{2}f_l \right) f_l\,,\\
\label{eq:REw5}
\frac{{\rm d}(w_5)_l}{{\rm d}l}&= \left(-4v_l w_l X'_l + f_l X_l - \frac{1}{2}f_l \right) (w_5)_l\,,\\
\label{eq:REw}
\frac{{\rm d}w_l}{{\rm d}l}&=\left(-4v_l w_l X'_l + f_l X_l + 4 v_l \right)w_l\,,\\
\label{eq:REh}
\frac{{\rm d}h_l}{{\rm d}l}&= \left(1- \frac{f_l}{4} \right)h_l\,,
\end{align}
from which we find four possible nontrivial fixed-point values of $f,\, w_5,\,w,\,h$:%
\footnote{The trivial fixed point, $f_\infty=(w_5)_\infty=w_\infty=h_\infty=0$, is stable only for $\epsilon<0$, and is not considered here.}
\begin{align}
\label{eq:FPVs}
({\rm i})&\quad f_\infty=\epsilon,\quad (w_5)_\infty=1,\quad w_\infty=h_\infty=0;\\
\label{eq:FPVs2}
({\rm ii})&\quad f_\infty=\frac{2}{3}\epsilon,\quad (w_5)_\infty= w_\infty= h_\infty=0;\\
\label{eq:FPVs3}
({\rm iii})&\quad f_\infty=\epsilon,\quad (w_5)_\infty=\frac{3}{7},\quad w_\infty= h_\infty=\infty;\\
\label{eq:FPVs4}
({\rm iv})&\quad f_\infty=2\epsilon,\quad (w_5)_\infty=0,\quad w_\infty= h_\infty=\infty.
\end{align}

Some remarks on the fixed points above are in order here. 
Since the magnetic field is external ($B\neq 0$), the fixed points (i) and (ii) with $h_\infty=0$ should be interpreted as corresponding to $C=0$.
We should note that the fixed points (iii) and (iv) are not usual in that the factors $X$ and $wX'$ in the RG equations (\ref{eq:REf})--(\ref{eq:REw}) are non-uniform in the limits $ w\rightarrow \infty$ and $h^2 \rightarrow \infty$: if one takes $w\rightarrow \infty$ first for fixed $h^2$, the fixed point (iii) is obtained; if one takes $h^2 \rightarrow \infty$ first for fixed $w$, the fixed point (iv) is obtained. 
In other words, the fixed point (iii) corresponds to the case $w_\infty \gg h_\infty^2 \gg1$, and the fixed point (iv) corresponds to the case $h_\infty^2  \gg w_\infty \gg1$.
The competition between $w\rightarrow\infty$ and $h^2 \rightarrow \infty$ in eq.~\eqref{eq:Paras2} can be characterized by introducing a dimensionless parameter,
\begin{align}
 \frac{h^2}{w}=\frac{C^2B^2}{\lambda_5\Gamma\chi\Lambda^2}.
 \label{eq:h2/w}
\end{align}
Then, we can see which parameters among $C,\lambda$ and $\lambda_5$ become dominant for a finite kinetic coefficient of the order parameter, $\Gamma$, and finite static susceptibilities, $\chi$ and $\chi_5$.
(Indeed, one can confirm that $\Gamma, \chi,$ and $\chi_5$ are finite by putting back the fixed-point values of $v, f, w_5, w$, and $h$ with eqs.~(\ref{eq:c5}) and (\ref{eq:c}) to the recursion relations (\ref{eq:RRchi}), (\ref{eq:RRchi5}), and (\ref{eq:RRGamma2}).)
By looking at $f_\infty$, $(w_5)_\infty$, and the fixed-point value of \eqref{eq:h2/w}, one can see the fixed point (iii) corresponds to $C\rightarrow0$ and $\lambda\rightarrow0$ with finite $\lambda_5$ (where the CME can be neglected compared to the diffusion effect), and the fixed point (iv) to $C\rightarrow\infty$, $\lambda\rightarrow 0$, and $\lambda_5\rightarrow\infty$ with $C^2/\lambda_5\rightarrow\infty$ (where the diffusion effect can be neglected compared to the CME). In short, we can regard the competition between the two limits $w \to \infty$ and $h^2 \to \infty$ as the competition between CME and diffusion.

\subsubsection{Stability of fixed points}
\label{sec:stability}
\begin{figure}[htb]
 \centering
 \includegraphics[bb= 0 0 810 768,  width=7.7cm]{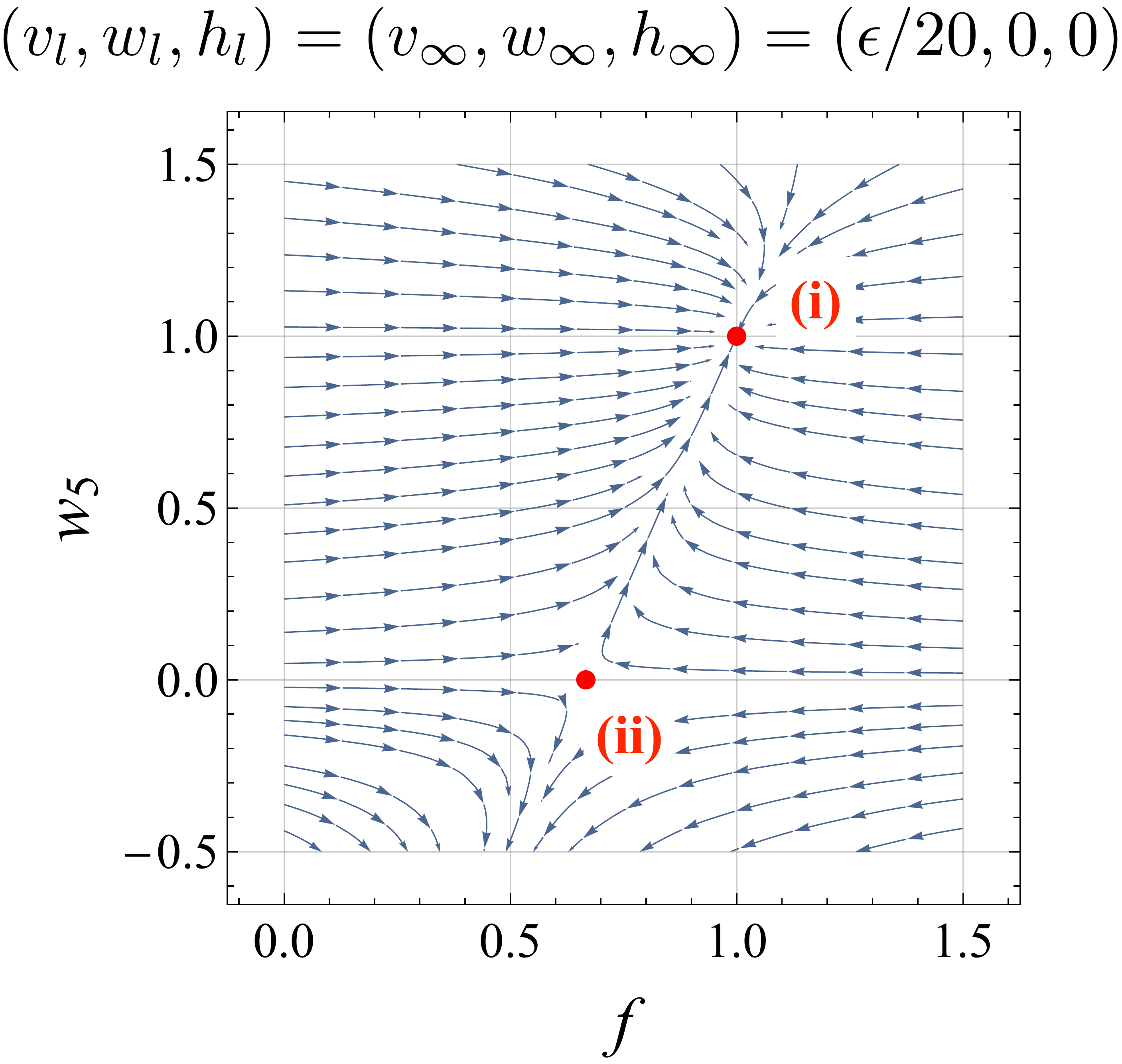}\\
 \caption{RG flow of the parameters $f$ and $w_5$ for fixed $w$ and $h$ ($\epsilon = 1$), which shows the fixed points (i) and (ii). }
 \label{fig:flow1}
\end{figure}

We first study the stabilities of the fixed points (i) and (ii). For this purpose, we consider the linear perturbations around the fixed points, 
\begin{align}
f_l=f_\infty+\delta f,\quad (w_5)_l=(w_5)_\infty+\delta w_5,\quad w_l=\delta w,\quad h_l=\delta h.
\end{align}
Substituting these expressions into eqs.~(\ref{eq:REf})--(\ref{eq:REh}) and setting $v_l=v_\infty = \epsilon/20$ from eq.~\eqref{eq:FVv},%
\footnote {Here we can ignore the fluctuation of $v_l$, because all of $v_l$ are multiplied by $O(\delta w)$ in eqs.~(\ref{eq:REf})--(\ref{eq:REh}).}  
the linearized equations with respect to $\delta f$, $\delta w_5$, $\delta w_5$, and $\delta h$ read
\begin{align}
\label{eq:linearized}
\frac{\dd}{\dd l}
\left(\begin{array}{c}
\delta f\\
\delta w_5\\
\delta w\\
\delta h
\end{array}\right)
=\left(\begin{array}{cccc}
-f_\infty\left(\theta_\infty+\dis \frac{1}{2}\right)&f_\infty^2 \theta_\infty^2&4 v_\infty f_\infty&0\\
(w_5)_\infty\left(\theta_\infty-\dis \frac{1}{2}\right)&\ f_\infty\left[\theta_\infty -\dis \frac{1}{2}-(w_5)_\infty\theta_\infty^2\right]&-4v_\infty(w_5)_\infty&0\\
0&0&f_\infty\theta_\infty+4v_\infty&0\\
0&0&0&1-\dis \frac{f_\infty}{4}
\end{array}\right)\left(\begin{array}{c}
\delta f\\
\delta w_5\\
\delta w\\
\delta h
\end{array}\right),
\end{align}
where we have defined 
\begin{align}
\label{eq:FPvs1and2}
\theta_\infty\equiv \frac{1}{1+(w_5)_\infty}=
 \left\{ 
 \begin{array}{c}
  \dis{\frac{1}{2}} \quad  {\rm for \ the \ case} \ ({\rm i}), 
   \vspace{5pt} \\
   1 \quad  \ {\rm for \ the \ case} \ ({\rm ii}).
 \end{array}
  \right. 
\end{align}
Because of $(w_5)_\infty\left(\theta_\infty-1/2\right)=0$ 
for both cases (i) and (ii), the $4 \times 4$ matrix in eq.~(\ref{eq:linearized}) 
is reduced to an upper triangular matrix. Then, the eigenvalues of the matrix are given just by its diagonal components for each fixed point:
\begin{align}
\label{eq:eigenvalues}
({\rm i})&\quad\left(-\epsilon,\,-\frac{\epsilon}{4},\,\frac{7}{10}\epsilon,\,1-\frac{\epsilon}{4} \right)\quad{\rm and }\quad({\rm ii})\quad\left(-\epsilon,\,\frac{\epsilon}{3},\,\frac{13}{15}\epsilon,\,1-\frac{\epsilon}{6} \right).
\end{align}
From this result, we find that the fixed point (ii) is unstable in the $w_5$ direction, and that the RG flow runs to the fixed point (i) 
(see also figure \ref{fig:flow1}
showing the RG flows in the $(f,w_5)$ plane at $w=h=0$). 
We also find that both fixed points are unstable in the directions of $w$ and $h$, showing that $\lambda$ and $C$ are relevant, so that small but nonzero values of $w$ and $h$ grow around the fixed point (i). 

\begin{figure}[tb]
\includegraphics[bb= 0 0 700 768, width=1\linewidth]{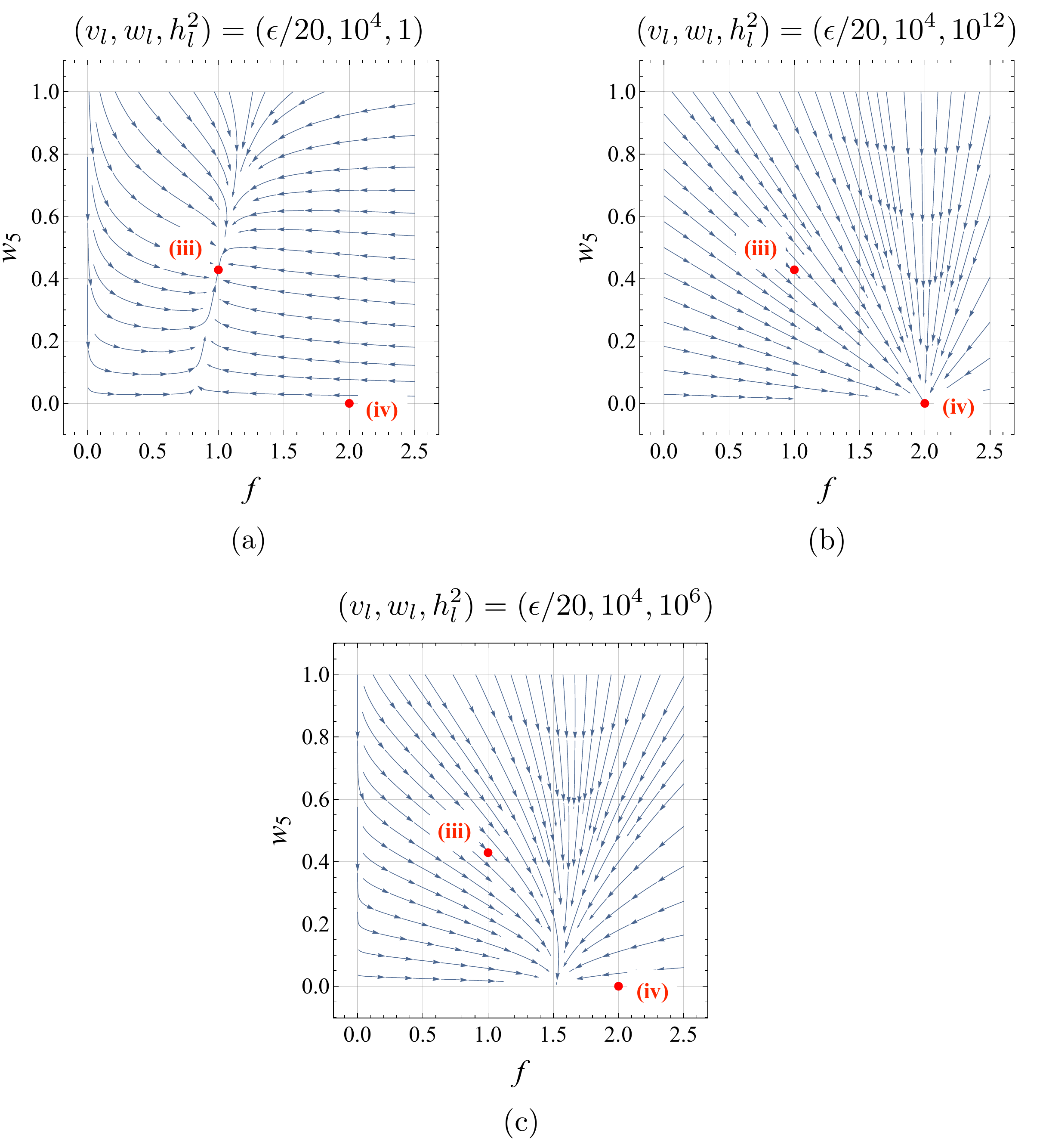}
\caption{RG flows of the parameters $f$ and $w_5$ ($\epsilon = 1$) with fixed values of $w$ and $h^2 $ in the cases (a) $w\gg h^2 \gg1$, (b) $h^2 \gg w\gg1$, and (c) $w\sim h^2 $. These figures show the existence of the fixed points (iii) and (iv), and the flow to their intermediate values.} \label{fig:flow2}
\end{figure}

\begin{figure}[htb]
 \centering
 \includegraphics[bb=0 0 730 768, 
 width=6.8cm]{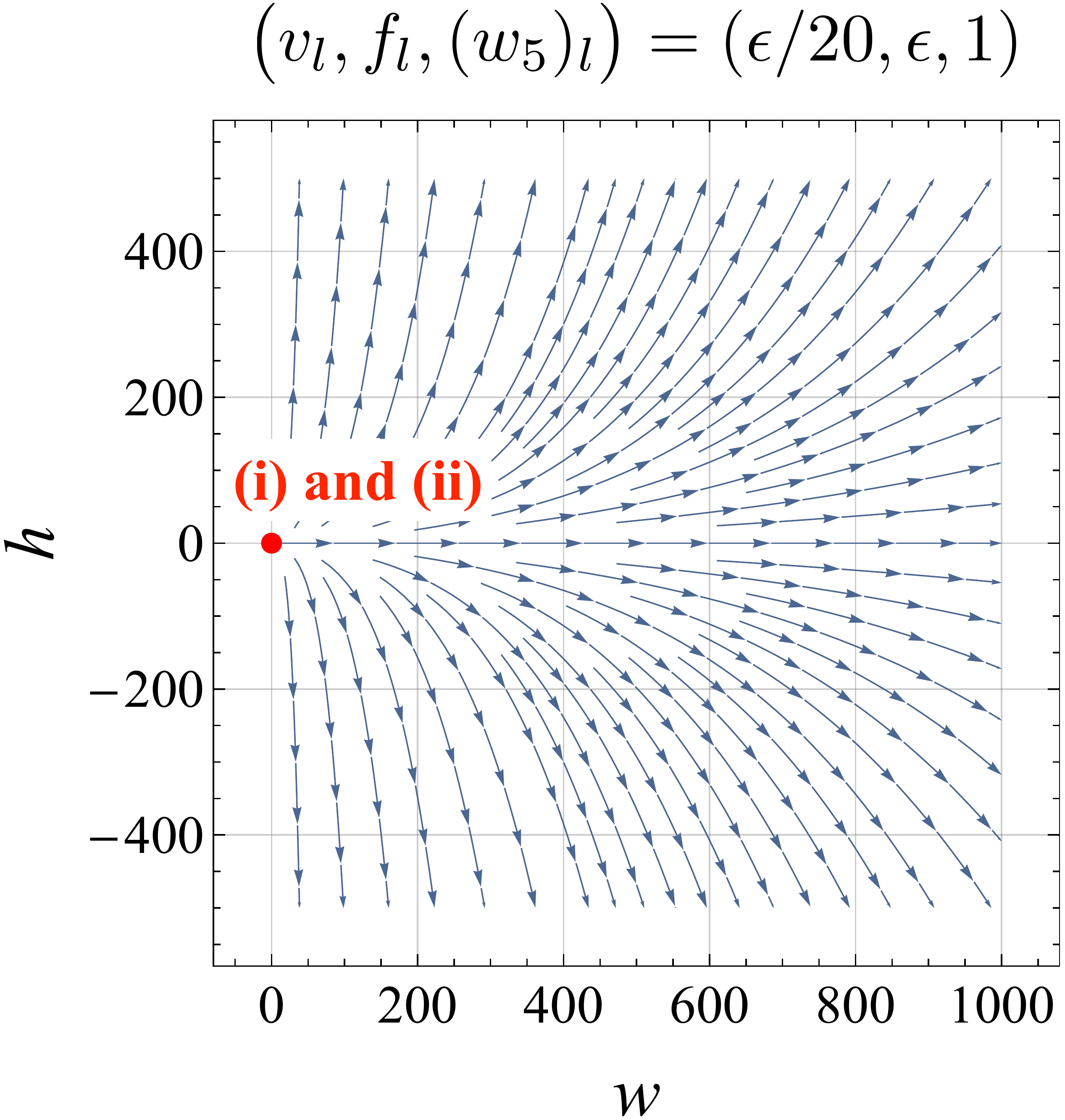}
 \caption{ RG flow of the parameters $w$ and $h$ with fixed values of 
 $f$ and $w_5$ ($\epsilon = 1$).
 }
 \label{fig:flow3}
\end{figure}

We shall first qualitatively discuss the RG flows by using flow diagrams. For this purpose, we here forcibly {\it fix} $w$ and $h$ to some finite values and investigate the RG flows in the $(f,w_5)$ plane. As we noted in the previous subsection, $f$ and $w_5$ flow to the fixed point (iii) when $w\gg h^2 \gg1$ while they flow to the fixed point (iv) when $h^2 \gg w\gg1$ 
(see the RG flows in figures \ref{fig:flow2}(a) and \ref{fig:flow2}(b), 
respectively). On the other hand, when $w \sim h^2$, they flow to the intermediate values between the fixed-point values of
(iii) and (iv), as shown in figure \ref{fig:flow2}(c).
We then vary $w$ and $h$ following the RG equations for fixed $f$ and $w_5$ close to the fixed point (i) or (ii).
As one can see in figure~\ref{fig:flow3}, unless $w \gg h$, the points in the $(w,h)$ plane flow in the direction along the $h$ axis. Therefore,  for most of the parameter region around the fixed point (i) or (ii), the system eventually flows to the fixed point (iv).

Next, we consider the RG flows in all the parameter space $(f,w_5,w,h)$. Here, we first set the initial parameters near the fixed point (i) and consider the flow equations at a fixed flow time. Similarly to the RG flows in the previous paragraph, when the initial values of $w$ and $h$ are varied, all the parameters move between the fixed-point values of (iii) and (iv).
The initial parameter region that flows to the fixed point (iii) is much broader than the region that flows to the fixed point (iv) in the $(w,h)$ plane, as is shown analytically within the linear-stability analysis around the fixed point (i) in appendix \ref{sec:crossover} (where a crossover between the dynamic universality classes corresponding to the fixed points (iii) and (iv) is also discussed). Therefore, in the almost whole region of the $(w, h)$ plane near the fixed point (i), the fixed point (iii) is unstable while the fixed point (iv) becomes stable.

When we consider the RG flow from the initial values near the fixed point (iii), all the parameters will eventually take the fixed-point values of (iv).%
\footnote{This is not the case of the RG evolution starting from the parameters exactly at the fixed point (iii).}
 This is because $h$ grows much more rapidly than $w$ due to the additional scaling factor $b$ in the recursion relation~(\ref{eq:RRh}) for $h$, compared to the relation~(\ref{eq:RRw}) for $w$. From the above discussion, it follows that the fixed point (iv) is stable in the almost whole region at finite $w$ and $h$, while generally at a finite flow time there is a small parameter region that flows to the fixed point (iii).

\subsection{Physical consequences}
\subsubsection{Dynamic universality class}
\label{sec:DUC}
The dynamic critical exponents are found by returning back to the recursion relation~(\ref{eq:RRGamma2}) for $\Gamma_l$ in each case of the fixed-point values (i)--(iv). The fixed points (i) and (iii) have the dynamic critical exponent of model E, $z=d/2$. This dynamic universality class is generally determined only by two-component order parameter and {\it one} conserved density that are coupled thorough the Poisson brackets. 
In our case, the order parameter field $\phi_\alpha$ and the axial isospin density $n_5$ are essential, whereas the isospin density $n$ does not affect the dynamic universality class.%
\footnote{There is a nonzero Poisson bracket among $n_5$ and $\phi_\alpha$ in eq.~(\ref{eq:PBn5phi}), whereas there are no nontrivial Poisson brackets among $n$ and $\phi_\alpha$.} 

On the other hand, the fixed point (iv) has the dynamic critical exponent $z=2$. This exponent is the same as that of model A up to $O(\epsilon)$, which is the dynamic universality class determined only by non-conserved order parameters. Here, the internal-momentum loop dominated by the CMW (the wavy lines in Fig.~\ref{fig:Sigma}) is suppressed near the fixed point (iv), so that not only $n$ but also $n_5$ do not affect the dynamic universality class. Actually, one can confirm that the factors $X$ and $wX'$ stemming from Fig.~\ref{fig:Sigma} vanish. 

In summary, we find that the dynamic universality class in the presence of the CME, corresponding to the stable fixed point (iv), is that of model A (see table \ref{tab:DUC}). Strictly speaking, there is a small parameter region that leads to model E, even $\Bv \neq {\bm 0}$ and $C\neq 0 $. Nevertheless, such a region is small compared to the region that leads to model A. We note that the dynamic universality class in table \ref{tab:DUC} remains unchanged even when the isospin chemical potential $\mu$ is absent.%
\footnote{One can easily confirm the dynamic universality class at $\mu = 0$ from the recursion relations which are obtained by setting $v=0$ in eqs.~(\ref{eq:REf})--(\ref{eq:REh}).}

\subsubsection{Critical attenuation}
As a result of the static critical behavior, eq.~(\ref{eq:eps}), we find the critical attenuation of the CMW: when the second-order chiral phase transition is approached, the speed of the CMW tends to zero as
\begin{align}
\label{eq:CMWcs}
v_{\rm CMW}^2\equiv \frac{C^2 B^2}{\chi\chi_5} \sim \xi^{-\frac{\alpha}{\nu}},
\end{align}
where $v_{\rm CMW}$ is the speed of the CMW \cite{Kharzeev:2010gd} and the ratio $\alpha/\nu$ is given by (\ref{eq:alpha/nu}). This phenomenon is analogous to the critical attenuation of the speed of (first) sounds near the critical point associated with the liquid-gas phase transition \cite{Onuki:1997} and the superfluid transition of liquid $^4$He \cite{Pankert:1986}.

\section{Conclusion and discussion}
\label{sec:CD}
In this paper, we have studied the critical dynamics near the second-order chiral phase transition in massless two-flavor QCD under an external magnetic field. Our main results are summarized in table~\ref{tab:DUC}. We also found the critical attenuation of the CMW analogous to that of the (first) sound waves in the liquid-gas phase transition and superfluid transition. 

We now discuss the similarity of the critical attenuation between the CMW and the sound wave of the compressive fluids near the liquid-gas critical point. Let us first recall the critical attenuation of sounds near the critical point associated with the liquid-gas phase transition where the order parameter $\psi$ is a linear combination of the energy density $\varepsilon$ and the mass density $\rho$. In this case, the speed of sound, $c_s$, is attenuated with the correlation-length dependence \cite{Onuki:1997},
\begin{align}
\label{eq:cs}
c_s^2\equiv \left( \frac{\partial P}{\partial \rho} \right)_{\! S}
= \frac{T\left( \dis \frac{\partial P}{\partial T} \right)^{\! 2}_{\! \rho}}{\rho^2 C_V\dis\left(1- \frac{C_V}{C_P}  \right)} \sim \xi^{-\frac{\alpha}{\nu}},
\end{align} 
where $P$ and $S$ are the pressure and total entropy per unit mass of the fluids, respectively. We used thermodynamic relations and the fact that the specific heat with constant volume $C_V\equiv T (\partial S/ \partial T)_\rho$, and that with constant pressure $C_P\equiv T (\partial S/ \partial T)_P$, diverge near the critical point as $C_V\sim\xi^{\frac{\alpha}{\nu}}$ and $C_P\sim\xi^{\frac{\gamma}{\nu}}$, respectively. Here, the critical exponents $\nu$, $\alpha$, and $\gamma$ defined by eq.~(\ref{eq:eps}) and $\psi\sim\tau^\gamma$ are determined by the static universality class of the 3D Ising model, $\alpha \approx 0.1,\,\nu \approx 0.6,\,\gamma \approx 1.2$. To obtain the last expression of eq.~(\ref{eq:cs}), we used the approximation ${C_V}/{C_P}\ll 1$ near the critical point. 
Remarkably, eq.~(\ref{eq:cs}) takes exactly the same form as that of CMW which we obtained in eq.~(\ref{eq:CMWcs}), although the values of $\alpha$ and $\nu$ themselves are different. 

Although we have ignored the energy-momentum tensor in this paper, it is crucial to include the motion of plasmas to investigate whether the CVE affects the dynamic critical phenomena in QCD. In this context, it is interesting to clarify whether the CVE coefficient does receive renormalization or not at the second-order chiral phase transition.%
\footnote{At the second-order chiral phase transition, the conventional proof of the non-renormalization theorem for the CVE coefficient is known to break down \cite{Golkar:2012kb}.} Moreover, it would be interesting to study the case with dynamical electromagnetic fields, where the so-called chiral plasma instability \cite{Akamatsu:2013pjd} might affect the critical dynamics of QCD.

To make our analysis more realistic, one needs to extend it to the case with finite quark masses where the QCD critical point is expected to appear in the $T$-$\mu_{\rm B}$ plane. 
Since non-linear fluctuations are suppressed in the presence of the CME, the CMW carried by the isospin densities can possibly affect the critical dynamics in realistic heavy-ion collisions, where $\pi^0$ as well as the Poisson bracket (\ref{eq:PBn5phi}) needs to be taken into account. Furthermore, if the critical attenuation of the CMW persists even in  QCD with finite quark masses, it would provide another possible signature of the QCD critical point in heavy-ion collision experiments.

\acknowledgments
We thank Y.~Hidaka for useful discussions, and Y.~Fujitani, Y.~Minami, and M.~A.~Stephanov for useful conversations.
We also thank the anonymous referee for pointing out a crucial error in the previous version of the paper and for feedback that improved the presentation. M.~H. is supported by the Special Postdoctoral Researchers Program and iTHES/iTHEMS Project (iTHEMS STAMP working group) at RIKEN. N.~S. is supported by JSPS KAKENHI Grant No.~17J04047.
N.~Y. is supported by JSPS KAKENHI Grant No.~16K17703 and MEXT-Supported Program for the Strategic Research Foundation at Private Universities, ``Topological Science'' (Grant No.~S1511006).

\appendix

\section{Chiral symmetry and susceptibilities $\chi$ and $\chi_5$}
\label{sec:chi}
The isospin and axial isospin susceptibilities defined in eq.~(\ref{chi}) can be decomposed as 
\begin{align}
\chi_{ab} 
 &= \frac{\partial n_{\text R,a}}{\partial \mu^b_\text R} 
 + \frac{\partial n_{\text L,a}}{\partial \mu^b_\text L} 
 + \frac{\partial n_{\text R,a}}{\partial \mu^b_\text L} 
 + \frac{\partial n_{\text L,a}}{\partial \mu^b_\text R},\\
 \chi_{5,ab} 
 &= \frac{\partial n_{\text R,a}}{\partial \mu^b_\text R}
 + \frac{\partial n_{\text L,a}}{\partial \mu^b_\text L}
 - \frac{\partial n_{\text R,a}}{\partial \mu^b_\text L} 
 - \frac{\partial n_{\text L,a}}{\partial \mu^b_\text R}.
\end{align}
Hence, when the chirality-mixing terms vanish, ${\partial n_{\text R,a}}/{\partial \mu^b_\text L} = {\partial n_{\text L,a}}/{\partial \mu^b_\text R} = 0$, it follows that $\chi_{ab}=\chi_{5,ab}$. 

Let us first consider the case with $\Bv={\bm 0}$. When $\mu=\mu_5=0$ (as considered in ref.~\cite{Rajagopal:1992qz}) or $\mu=\mu_5\neq0$, the system respects chiral symmetry $\text{SU}(2)_\text L \times \text{SU}(2)_\text R$ or $\text{SU}(2)_\text L\times \text U(1)^3_\text {V+A}$, respectively (see table~\ref{tab:sym}). 
Because both ${\partial n_{\text R,a}}/{\partial \mu^b_\text L} = \langle n_{\text R,a} n_{\text L,b} \rangle$ and ${\partial n_{\text L,a}}/{\partial \mu^b_\text R} = \langle n_{\text L,a} n_{\text R,b} \rangle$ are {\it not} invariant under these {\it non-Abelian} chiral transformations, such mixing terms vanish, and, as a result, we have $\chi_{ab}=\chi_{5,ab}$.
On the other hand, when $\Bv\neq {\bm 0}$, the system respects the {\it Abelian} symmetry $\mathcal G$ alone (see table~\ref{tab:sym}). As $\langle n_{\text R,a} n_{\text L,b}\rangle$ and $\langle n_{\text L,a} n_{\text R,b} \rangle$ are invariant under the subgroup $\mathcal G$, the symmetry provides no constraint on ${\partial n_{\text R,a}}/{\partial \mu^b_\text L}$ and ${\partial n_{\text L,a}}/{\partial \mu^b_\text R}$, and hence, we have generally $\chi_{ab} \neq \chi_{5,ab}$. This justifies the reason why we set $\chi$ and $\chi_5$ as independent parameters in eq.~(\ref{eq:GL}). 

\section{Jacobian}
\label{sec:Jacobian}
In this appendix, we show that, if one keeps the Jacobian $J \equiv \det \left( \partial_t - \delta \mathcal F / \delta \psi \right)$ in eq.~(\ref{eq:identMSR}), it cancels the diagrams with the so-called {\it closed response loops} \cite{Tauber} that arise by the contractions of {\it internal} $\tilde\psi_M$ and $\psi_N$. Since we are considering Gaussian white noises, non-linear interactions can be derived only from the term proportional to $\tilde\psi_N\mathcal F_N$ in eq.~(\ref{eq:MSR}). This term produces the same diagrams containing closed response loops as the diagrams produced by the following effective interaction, 
\begin{align}
\label{eq:effectiveinteraction}
- \left< \tilde\psi_N \psi_M \right>_{\! {\rm free}}\frac{\delta \mathcal F_N}{\delta \psi_M}\propto- \theta(0)\frac{\delta \mathcal F_N}{\delta \psi_N}.
\end{align} 
This is obtained by writing $-\tilde \psi_N \mathcal F_N=-\tilde\psi_N \psi_M \delta \mathcal F_N/\delta \psi_M$ and replacing $\tilde\psi_N \psi_M$ by the closed response loop between $\tilde \psi _N$ and $\psi_M$.  To obtain the right-hand side of eq.~(\ref{eq:effectiveinteraction}), we have also used $\left< \tilde\psi_N(t) \psi_M(t') \right> _{\rm free} \propto \theta(t-t')$. On the other hand, one can write the Jacobian into the following form:
\begin{align}
\label{eq:det} 
J \propto \exp \ln \det \left[1 -   \left( \frac{\partial}{\partial t} \right)^{\! \! -1}\dis \frac{\delta \mathcal F_N}{\delta \psi_M} \right]=\exp \left\{-{\rm tr}\sum_{n=1}^{\infty}\left[\left(\frac{\partial}{\partial t}\right)^{\! \! -1}\dis \frac{\delta \mathcal F_N}{\delta \psi_M} \right]^n\right\}=\exp \left[-\theta(0)\dis \frac{\delta \mathcal F_N}{\delta \psi_N} \right].
\end{align}
Here, we have omitted a factor independent of the fields and used $\partial \theta(t-t')/\partial t=\delta(t-t')$ to obtain the inverse operator of time derivative. It can be readily checked that contributions for $n \geq 2$  in the summation vanish. Then, one finds that the effective action stemming from the functional determinant (\ref{eq:det}) cancels the effective interaction (\ref{eq:effectiveinteraction}).

\section{Calculation of the self-energies and the vertex function}
\label{sec:calculation}
In this appendix, we evaluate the self-energies of the order parameter, $\Sigma_{\alpha\beta}(\kv,\omega)$, and those of the conserved charge densities, $\Pi_{ij}(\kv,\omega)$, to the first order of $\epsilon$, namely at the one-loop level. From these expressions, we compute their derivatives with respect to $\omega$, $\kv$ or ${\bm k}^2$ that are used in eqs.~(\ref{eq:RRGamma})--(\ref{eq:RRCchi5}), (\ref{eq:ctildec}), and (\ref{eq:ctilde5c5}). Then we derive eqs.~(\ref{eq:RRGamma2})--(\ref{eq:RRC}). We also evaluate the three-point vertex function used in eq.~(\ref{eq:RRgchi5}) at the one-loop level, and derive eq.~(\ref{eq:RRg}).

\subsection{Self-energy $\Pi$}
Let us begin with $\Pi_{ij}(\kv,\omega)$, whose diagram is given by figure~\ref{fig:Pi},

\begin{figure}[b]
\centering\begin{tabular}{cc}
\begin{minipage}{.5\textwidth}
\centering
\includegraphics[bb=0 0 215 137,height=3cm]{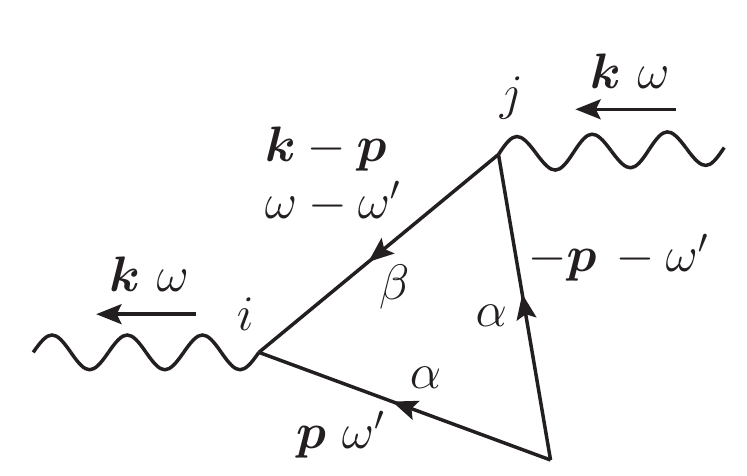}
\caption{Diagram for $\Pi_{ij}$ at one-loop level} 
\end{minipage}
\end{tabular}\\　\\
\centering\begin{tabular}{cc}
\begin{minipage}{.7\textwidth}
\centering
\subfigure[]{ \includegraphics[bb=0 0 215 122,height=2.8cm]{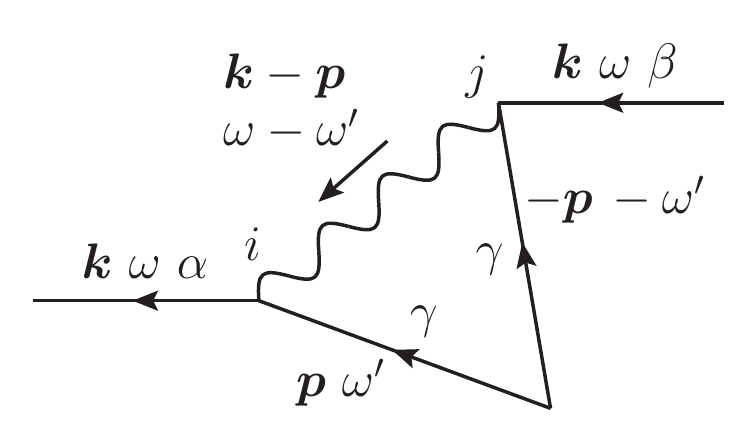}\label{fig:Sigma(a)}}~
\subfigure[]{ \includegraphics[bb=0 0 218 133,height=2.8cm]{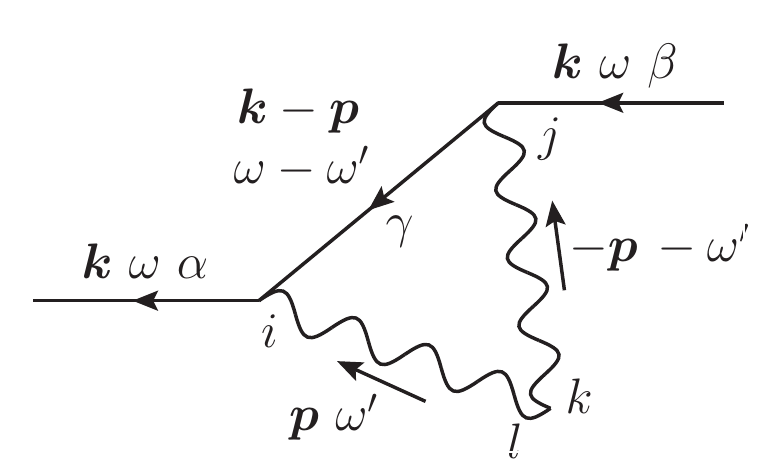}\label{fig:Sigma(b)}}
\caption{Diagrams for $\Sigma_{\alpha\beta}$ at one-loop level}\label{fig:Sigma}
\end{minipage}
\end{tabular}
\end{figure}

\begin{align}
\Pi_{ij}(\kv,\omega)
&=
\int \frac{\dd \omega'}{2\pi}\int \frac{\dd^d p}{(2\pi)^d}
V^0_{i; \alpha \beta}(\kv,\pv) G^0(\kv-\pv,\omega-\omega') |G^0(\pv,\omega')|^2 2\Gamma V^0_{\beta;\alpha j}
\notag\\
&=
\int \frac{\dd^d p}{(2\pi)^d}
\frac{ F _{ij}}
{\{-i\omega+\Gamma[r+(\kv-\pv)^2]+\Gamma(r+\pv^2)\}(r+\pv^2)}\,,
\end{align}
where we defined $F_{ij}$ as the components of the following matrix:
\begin{align}
F\equiv
\left(
\begin{array}{cc}
8\gamma^2\Gamma\lambda \kv^2&0\\
8i\gamma^2 C \Gamma \Bv\cdot\kv&-2g^2[(\kv-\pv)^2-\pv^2]/\chi_5
\end{array}
\right).
\end{align}
From this expression, we obtain,
\begin{align}
\left.  i \frac{\d \Pi_{21}(\bm{k},0)}{\d \bm{k}}    \right|_{\bm{k}\rightarrow \bm{0}}&=-4\gamma^2 C\Bv
\int \frac{\dd^d p}{(2\pi)^d}\left(
\frac{1}{\pv^4}+O(\epsilon)\right)=-\frac{\gamma^2 C\Bv \Lambda^{-\epsilon}}{2\pi^2} \ln b,
\label{eq:dPidk21}\\
\frac{1}{2}\left.  \frac{\d^2 \Pi_{11} (\bm{k},0)}{\d \bm{k}^2}    \right|_{\bm{k}\rightarrow \bm{0}}
&=4\gamma^2\lambda
\int \frac{\dd^d p}{(2\pi)^d}\left(
\frac{1}{\pv^4}+O(\epsilon)\right)
\label{eq:dPidk11}
=\frac{\gamma^2\lambda\Lambda^{-\epsilon}}{2\pi^2} \ln b,
\end{align}
where we carried out the integral over ${\bm p}$ in the shell $\Lambda/b<|\pv|<\Lambda$ using the standard formula in the dimensional regularization:
\begin{align}
\int \frac{\dd^d p}{(2\pi)^d}\frac{1}{\pv^4}=\frac{\Lambda^{-\epsilon}}{8\pi^2}\ln b.
\end{align}
We also have
\begin{align}
\frac{1}{2}\left.  \frac{\d^2 \Pi_{22} (\bm{k},0)}{\d \bm{k}^2}    \right|_{\bm{k}\rightarrow \bm{0}}
&=\frac{g^2}{\chi_5 \Gamma}
\int \frac{\dd^d p}{(2\pi)^d}\left(
\frac{\cos 2\theta}{\pv^4}+O(\epsilon)\right)
\label{eq:dPidk22}
=-\frac{g^2 \Lambda^{-\epsilon}}{16\pi^2 \chi_5 \Gamma} \ln b,
\end{align}
where we parameterized $\pv$ in the limit of $d\rightarrow 4$ by
\begin{align}
\label{eq:ppara}
p_1=p\cos\theta,\quad p_2=p\sin\theta\cos\phi,\quad p_3=p\sin\theta\sin\phi\cos\varphi,\quad p_4=p\sin\theta\sin\phi\sin\varphi
\end{align}
with $\kv=(1,0,0,0)$ and used
\begin{align}
\int \frac{\dd^d p}{(2\pi)^d}\frac{\cos 2 \theta}{\pv^4}=-\frac{\Lambda^{-\epsilon}}{16\pi^2} \ln b\,.
\end{align}

Because $\Pi_{ij}(\zev,\omega)=0$, eqs.~(\ref{eq:ctildec}) and (\ref{eq:ctilde5c5}) reduce to
\begin{align}
\label{eq:ctilde2}
\tilde{c}&=d-c,\quad
\tilde{c}_5=d-c_5.
\end{align}
Furthermore, by using eqs.~(\ref{eq:dPidk21}), (\ref{eq:dPidk11}), (\ref{eq:dPidk22}), and (\ref{eq:ctilde2}), eqs.~(\ref{eq:RRlambdachi})--(\ref{eq:RRCchi5}) provide the following recursion relations:
\begin{align}
\label{eq:RRlambdachi2}
\frac{\lambda_{l+1}}{\chi_{l+1}}&=b^{z-2}\frac{\lambda_l}{\chi_{l}}\left(1- 4v_l \ln b
  \right),\\
\label{eq:RRlambda5chi52}
\frac{(\lambda_5)_{l+1}}{(\chi_5)_{l+1}}&=b^{z-2}
\frac{(\lambda_5)_l}{(\chi_5)_l}
\left(1+\frac{f_l}{2}\ln b \right),\\
\label{eq:RRCchi2}
\frac{C_{l+1} }{\chi_{l+1}}&=b^{z+c_5-c-1}\frac{C_l }{\chi_l}\left(1-4v_l \ln b \right),\\
\frac{C_{l+1} }{(\chi_5)_{l+1}}&=b^{z+c-c_5-1}\frac{C_l }{(\chi_5)_l}.
\end{align}
Then, we can derive eqs.~(\ref{eq:RRlambda2})--(\ref{eq:RRC}) by using eqs.~(\ref{eq:RRchi}), (\ref{eq:RRchi5}), and (\ref{eq:RRlambdachi2})--(\ref{eq:RRCchi2}).

\subsection{Self-energy $\Sigma$}
Next, let us evaluate the self-energy $\Sigma_{\alpha\beta}(\kv,\omega)$, 
whose diagrams are 
shown in figure~\ref{fig:Sigma},
\begin{align}
\Sigma_{\alpha\beta}(\kv,\omega)&=\Sigma^{(a)}_{\alpha\beta}(\kv,\omega)+\Sigma^{(b)}_{\alpha\beta}(\kv,\omega),
\end{align} 
where $\Sigma^{(a)}$ and $\Sigma^{(b)}$ correspond to figures~\ref{fig:Sigma(a)} and \ref{fig:Sigma(b)}, respectively. 
Their diagonal  components used in eqs.~(\ref{eq:RRGamma}) and (\ref{eq:atildea}) are given by
\begin{align}
\Sigma_{\alpha\alpha}^{(a)}(\kv,\omega)&=\int \frac{\dd \omega'}{2\pi}\int \frac{\dd^d p}{(2\pi)^d}
V^0_{\alpha; \gamma i} D^0_{ij}(\kv-\pv,\omega-\omega') |G^0(\pv,\omega')|^2 2\Gamma V_{j;\gamma\alpha} (\kv-\pv,-\pv)\notag\\
&=
4\gamma^2\Gamma\int \frac{\dd^d p}{(2\pi)^d}
\frac{
 \lambda(\kv-\pv)^2[-i \omega +\Gamma(r+\pv)^2 +\lambda_5(\kv-\pv)^2/\chi_5 ]+  C^2 [\Bv\cdot(\kv-\pv)]^2/\chi_5  }
{(r+\pv^2)\det [D ^0(\kv-\pv,\omega+i\Gamma(r+\pv^2))]^{-1}}\notag\\
&\quad+\frac{g^2}{\chi_5}\int \frac{\dd^d p}{(2\pi)^d}
\frac{(\pv^2-\kv^2)[-i\omega+ \Gamma(r+\pv)^2 +\lambda(\kv-\pv)^2/\chi]
}
{(r+\pv^2)\det [D ^0(\kv-\pv,\omega+i\Gamma(r+\pv^2))]^{-1}}\,,
\label{eq:Sigma(a)2} \\
\Sigma_{\alpha\alpha}^{(b)}(\kv,\omega)
&=\int \frac{\dd \omega'}{2\pi}\int \frac{\dd^d p}{(2\pi)^d}
V^0_{\alpha; \gamma i}G^0(\kv-\pv,\omega-\omega') B^0 _{ij}(\pv,\omega') V_{\gamma;\alpha j}\notag\\
&=8\gamma^2\Gamma^2
\int \frac{\dd \omega'}{2\pi}\int \frac{\dd^d p}{(2\pi)^d}
\frac{\lambda\pv^2(\omega'^2+\kappa_5\pv^2)+\lambda_5 (C\Bv\cdot\pv)^2/\chi_5^2 }{|\det [D^0(\pv,\omega')]^{-1}|^2\{-i(\omega-\omega')+\Gamma[r+(\kv-\pv)^2]\}}\notag\\
&\quad - \frac{2g^2}{\chi_5^2}\int \frac{\dd \omega'}{2\pi}\int \frac{\dd^d p}{(2\pi)^d}
\frac{ \lambda_5 \pv^2(\omega'^2+\kappa\pv^2)+\lambda (C\Bv\cdot\pv)^2/\chi^2}{|\det [D^0(\pv,\omega')]^{-1}|^2\{-i(\omega-\omega')+\Gamma[r+(\kv-\pv)^2]\}}\,.
\label{eq:Sigma(b)b}
\end{align}
Note that we do not take a sum over $\alpha$ here.

The integral over $\omega'$ in eq.~(\ref{eq:Sigma(b)b}) can be performed by closing contour below and picking up the poles on a lower half plane,
\begin{align}
\det [D^0(\pv,\omega')]^{-1}
&=\left[-i\omega'+\kappa_+ \pv^2+ i\Omega(\pv)\right]\left[-i\omega'+\kappa_+ \pv^2- i\Omega(\pv)\right]=0,
\label{eq:detM-1}
\end{align}
where we defined
\begin{align}
\Omega(\pv)\equiv\sqrt{\frac{(C\Bv\cdot\pv)^2}{\chi\chi_5} - \kappa_-^2  \pv^4 },\quad\kappa_\pm\equiv\frac{1}{2}\left(\frac{\lambda}{\chi}\pm\frac{\lambda_5}{\chi_5}\right).
\end{align}
The result of the contour integral is given by
\begin{align}   
\Sigma_{\alpha\alpha}^{(b)}(\kv,\omega)&=-2\gamma^2\Gamma^2 \int \frac{\dd^d p}{(2\pi)^d} \left(
\frac{ - \lambda \kappa_- \pv^4 +  (C\Bv\cdot\pv)^2/\chi_5  }
{i\Omega(\pv)}K_-(\pv,\kv)-\lambda\pv^2K_+(\pv,\kv)\right)\notag\\
&\quad
+\frac{ g^2}{2\chi_5^2}
 \int \frac{\dd^d p}{(2\pi)^d} \left(
\frac{ \lambda_5 \kappa_- \pv^4 + (C\Bv\cdot\pv)^2/\chi  }
{i\Omega(\pv)}K_-(\pv,\kv)-\lambda_5 \pv^2K_+(\pv,\kv)
\right),
\label{eq:Sigma(b)2}
\end{align}
where we introduced
\begin{align}
\label{eq:K} 
K_{\pm}(\pv,\kv)&\equiv
\frac{1}{ [ \kappa_+ \pv^2 + i\Omega(\pv)] [-i\omega+\Delta_+(\pv,\kv)+ i\Omega(\pv)]}
\pm
\frac{1}{ [ \kappa_+ \pv^2 - i\Omega(\pv)] [-i\omega+\Delta_+(\pv,\kv)- i\Omega(\pv)]},\\
\Delta_+(\pv,\kv)&\equiv\Gamma[r+(\kv-\pv)^2]+\kappa_+\pv^2.
\end{align}
Using an identity,
\begin{align}
 \det [D^0 (\pv,\omega+i\Gamma[r+(\kv-\pv)^2])]^{-1} = [-i\omega+\Delta_+(\pv,\kv)]^2+ \Omega(\pv)^2,
\end{align}
together with some straightforward calculations, we obtain
\begin{align}
\Sigma_{\alpha\alpha}^{(b)}(\kv,\omega)
&=4\gamma^2\chi\Gamma^2 \int \frac{\dd^d p}{(2\pi)^d} 
\dis\frac{-i\omega+\lambda_5\pv^2/\chi_5+\Gamma[r+(\kv-\pv)^2]}{  \det [D^0 (\pv,\omega+i\Gamma[r+(\kv-\pv)^2])]^{-1}} \nonumber \\
& \quad -\frac{ g^2}{\chi_5} \int \frac{\dd^d p}{(2\pi)^d} 
\dis\frac{-i\omega+\lambda \pv^2/\chi+\Gamma[r+(\kv-\pv)^2]}{  \det [D^0 (\pv,\omega+i\Gamma[r+(\kv-\pv)^2])]^{-1}  }.
\label{eq:Sigma(b)4}
\end{align}

Combining eqs.~(\ref{eq:Sigma(a)2}) and (\ref{eq:Sigma(b)4}), we find 
\begin{align}
\Sigma_{\alpha\alpha}(\kv,\omega)
&=
4\gamma^2\chi\Gamma\int \frac{\dd^d p}{(2\pi)^d}
\frac{
 \Theta(\pv,\kv,0)\Theta_5(\pv,\kv,\omega)+   [C \Bv\cdot(\kv-\pv)]^2/(\chi\chi_5)  }
{(r+\pv^2)[ \Theta(\pv,\kv,\omega)\Theta_5(\pv,\kv,\omega) + (C\Bv\cdot\pv)^2 /(\chi\chi_5)  ]}\notag\\
&\quad -\frac{g^2}{\chi_5}\int \frac{\dd^d p}{(2\pi)^d}
\frac{(r+\kv^2)\Theta(\pv,\kv,\omega)
}
{(r+\pv^2)[ \Theta(\pv,\kv,\omega)\Theta_5(\pv,\kv,\omega)+ (C\Bv\cdot\pv)^2 /(\chi\chi_5)  ]},
\label{eq:Sigma_alal}
\end{align}
where $\Theta(\pv,\kv,\omega)$ and $\Theta_5(\pv,\kv,\omega)$ are defined by
\begin{align}
\Theta(\pv,\kv,\omega)&\equiv-i\omega+\Gamma(r+\pv^2)+\lambda(\kv-\pv)^2/\chi, \\
\Theta_5(\pv,\kv,\omega)&\equiv-i\omega+\Gamma(r+\pv^2)+\lambda_5(\kv-\pv)^2/\chi_5 .
\end{align}
Note that $\Sigma_{11}=\Sigma_{22}$ and the first term on the right-hand side of eq.~(\ref{eq:Sigma_alal}) is independent of $\kv$ when $\omega=0$. From the expression~(\ref{eq:Sigma_alal}), we obtain
\begin{align}
\label{eq:dSigdk2mu}
\frac{1}{2} \left.   \frac{\d^2 \Sigma_{\alpha\alpha}(\bm{k},0)}{\d \bm{k}^2}    \right|_{\bm{k}\rightarrow \bm{0}}&=-\frac{g^2}{\chi_5} \int \frac{\dd^d p}{(2\pi)^d}  \left( \frac{ \Theta(\pv,\zev,0)}{(r+\pv^2)[\Theta(\pv,\zev,0)\Theta_5(\pv,\zev,0)+ (C\Bv\cdot\pv)^2 /(\chi\chi_5) ] }
 + O(\epsilon)  \right)
,\\
i \left.   \frac{\d \Sigma_{\alpha\alpha}(\bm{0},\omega)}{\d \omega}   \right|_{\omega\rightarrow0}
&=
-4\gamma^2\chi\Gamma \int \frac{\dd^d p}{(2\pi)^d}
\frac{\Theta_5(\pv,\zev,0)}{(r+\pv^2)\{\Theta(\pv,\zev,0)\Theta_5(\pv,\zev,0)+ (C\Bv\cdot\pv)^2 /(\chi\chi_5) \} }\notag\\
&\quad+\frac{g^2}{\chi_5}
\int \frac{\dd^d p }{(2\pi)^d}
\frac{r[ \Theta^2(\pv,\zev,0)-(C\Bv\cdot\pv)^2/(\chi\chi_5)^2 ]}
{(r+\pv^2) [\Theta(\pv,\zev,0)\Theta_5(\pv,\zev,0) +(C\Bv\cdot\pv)^2/(\chi\chi_5)^2   ]^2}.
\label{eq:Sigmaomega}
\end{align}
The $O(\epsilon)$ terms in eq.~\eqref{eq:dSigdk2mu} proportional to $r$ are irrelevant for the following discussion. 
Let us carry out the integral over $\pv$ in the shell $\Lambda/b<|\pv|<\Lambda$. 
Setting $\Bv=(1,0,0,0)$ and using the parameterization~(\ref{eq:ppara}), we obtain
\begin{align}
\frac{1}{2\Gamma} \left.   \frac{\d^2 \Sigma_{\alpha\alpha}(\bm{k},0)}{\d \bm{k}^2}    \right|_{\bm{k}\rightarrow \bm{0}}
&=-\frac{2f}{\sqrt{(1+w)(1+w_5)+h^2}+\sqrt{(1+w)(1+w_5)}}\sqrt{\frac{1+w}{1+w_5}}\ln b.
\label{eq:Sigmakk}
\end{align}
where $w,\,w_5,\,h,\,f$ are defined in eq.~(\ref{eq:Paras}). 
Here, we have carried out the integral over $\theta$ by using the 
following formula (with $a$ being a real constant):
\begin{align}
\label{eq:INTtheta}
\int_0^\pi \dd\theta
\frac{\sin^2\theta}{a^2+\cos^2\theta}
=\frac{\pi}{a(\sqrt{a^2+1}+a^2)}.
\end{align}
To the leading order of $\epsilon$, we can ignore the term proportional to $r$ on the right-hand side of (\ref{eq:Sigmaomega}). 
Then, by comparing eqs.~(\ref{eq:dSigdk2mu}) and (\ref{eq:Sigmaomega}), we can rewrite eq.~(\ref{eq:Sigmaomega}) as
\begin{align}
i \left.   \frac{\d \Sigma_{\alpha\alpha}(\bm{0},\omega)}{\d \omega}   \right|_{\omega\rightarrow0}
&=-\frac{8vw}{\sqrt{(1+w)(1+w_5)+h^2}+\sqrt{(1+w)(1+w_5)}}\sqrt{\frac{1+w_5}{1+w}}\ln b,
\label{eq:Sigmaomega2}
\end{align}
where $v$ is defined in eq.~(\ref{eq:Parastas}). Using eqs.~(\ref{eq:Sigmakk}) and (\ref{eq:Sigmaomega2}), 
eqs.~(\ref{eq:RRGamma}) and (\ref{eq:atildea}) become
\begin{gather}
\label{eq:recurionGamma2}
\Gamma_{l+1}=\Gamma_{l}
\left(1+f_l X_l \ln b
\right)b^{d+z-\tilde a -a-2}, \\
\label{eq:atilde2}
1=\left(  1+4 v_l w_l  X'_l \ln b \right)b^{d-\tilde a -a},
\end{gather}
respectively, where $X$ and $X'$ are defined in eq.~(\ref{eq:Paras2}). 
By substituting eq.~(\ref{eq:atilde2}) into eq.~(\ref{eq:recurionGamma2}), we finally arrive at eq.~(\ref{eq:RRGamma2}).

\subsection{Vertex function $\mathcal V$}

\begin{figure}[t]
\centering
\begin{minipage}[c]{0.75\hsize}
\subfigure[]{\includegraphics[bb=0 0 215 122,height=3cm]{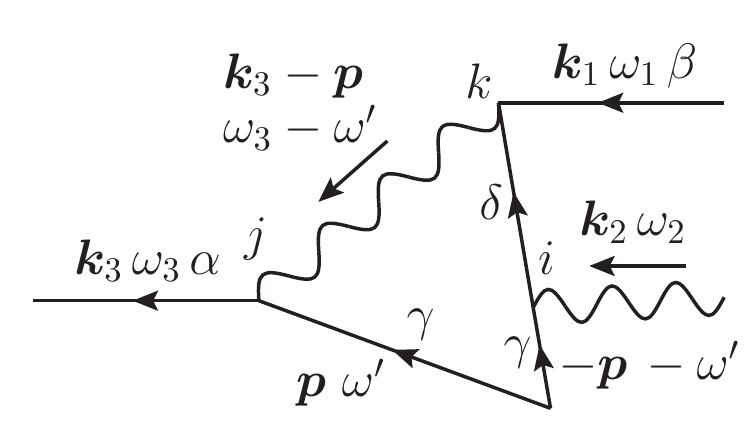}\label{Fig:phina}}~
\subfigure[]{ \includegraphics[bb=0 0 217 127,height=3cm]{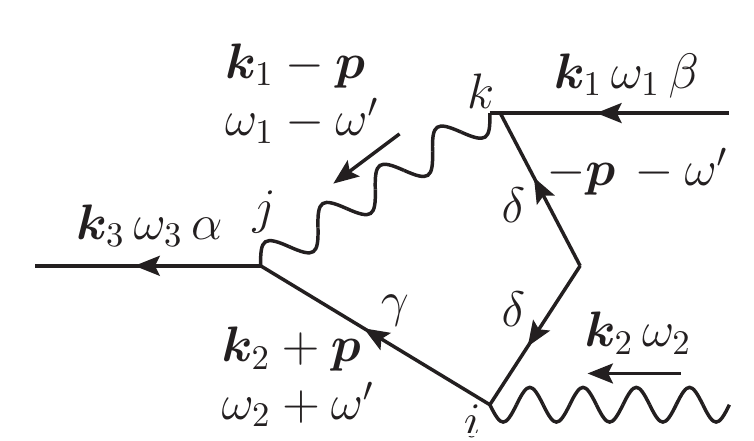}\label{Fig:phinb}}~
\end{minipage}\\
\begin{minipage}[c]{0.75\hsize}
\subfigure[]{\includegraphics[bb=0 0 222 125,height=3cm]{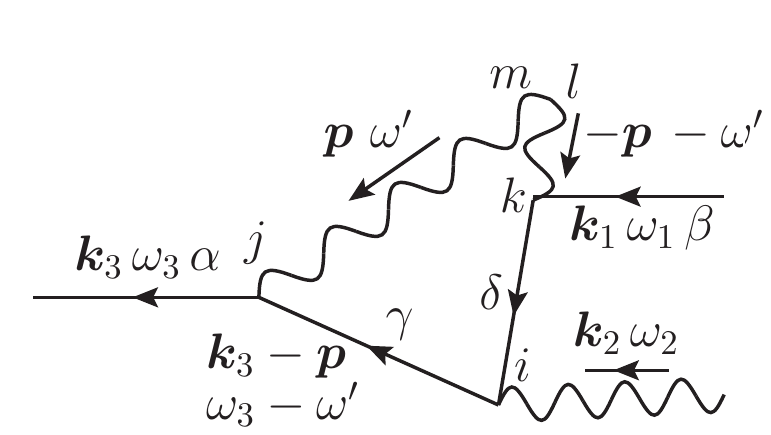}\label{Fig:phinc}}~
\subfigure[]{\includegraphics[bb=0 0 215 122,height=3cm]{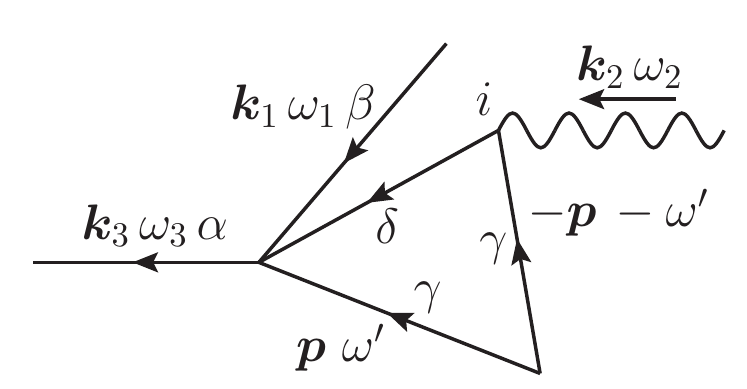}\label{Fig:phind}}
\end{minipage}
\caption{Diagrams for $V_{\alpha;\beta i}$ at one-loop level}\label{Fig:VC}
\end{figure}

The lowest-order diagrams for $\mathcal V_{\alpha;\beta i}$ are depicted in figure~\ref{Fig:VC}. We find that these diagrams with $\alpha\neq\beta$ and $i=2$ satisfy the following identity in the limit $\kv_1,\kv_2\rightarrow\zev$ and $\omega_1,\omega_2\rightarrow0$:
\begin{align}
\label{eq:V}
 \mathcal V_{\alpha;\beta2} (\zev,\zev,0,0)  = -i \frac{g \varepsilon_{\alpha\beta}}{\chi_5}    \left.   \frac{\d \Sigma_{\gamma\gamma}(\bm{0},\omega)}{\d \omega}   \right|_{\omega\rightarrow0}.
\end{align}
Note that the contribution from figure~\ref{Fig:phind} vanishes as long as $\alpha\neq\beta$. By substituting eqs.~(\ref{eq:V}), (\ref{eq:RRchi5}), and (\ref{eq:atildea}) into eq.~(\ref{eq:RRgchi5}), we obtain eq.~(\ref{eq:RRg}). 

We can derive eq.~(\ref{eq:V}) as a consequence of the Ward-Takahashi identity as follows. Our derivation is similar to that of ref.~\cite{ONWT} (see its section~III.\,A), where the Ward-Takahashi identity for O($N$) symmetric systems is derived. We shall begin with the following generating functional,
\begin{align}
Z[\tilde \jmath, j, \tilde \mu_5,\mu_5]
 \equiv 
 \left<  \exp \int \dd t \int \dd \rv \left(  \tilde \jmath _\alpha \tilde \phi_\alpha + j_\alpha\phi_\alpha + \tilde \mu_5 \tilde n_5 + \mu_5 n_5  \right)   \right>.
\end{align}
Here, $\tilde \jmath _\alpha,\, j_\alpha,\,\tilde \mu_5,$ and $\mu_5$ are the external fields of 
$\tilde \phi_\alpha, \,\phi_\alpha,\, \tilde n_5$ and 
$n_5$, 
respectively. We define
\begin{align}
\label{eq:Phi}
\tilde \Phi_\alpha \equiv \frac{\delta \ln Z}{\delta \tilde \jmath_\alpha},\quad \Phi_\alpha \equiv \frac{\delta \ln Z}{\delta j_\alpha},\quad
\tilde N_5 \equiv \frac{\delta \ln Z}{\delta \tilde \mu_5},\quad N_5 \equiv \frac{\delta \ln Z}{\delta \mu_5},
\end{align}
which reduce to the expectation values of $\tilde \phi_\alpha, \,\phi_\alpha,\, \tilde n_5$ and $n_5$, respectively, when the external fields are set to zero. 
We also define the following effective actions by 
the Legendre transformations of eq.~(\ref{eq:Phi}):
\begin{align}
\label{eq:W}
W[\tilde \Phi,\Phi,\tilde \mu_5,\mu_5]&\equiv-\ln Z[\tilde \jmath,j, \tilde \mu_5,\mu_5] +   \int \dd t \int \dd \rv \left(\tilde \jmath_\alpha \tilde \Phi_\alpha+j_\alpha\Phi_\alpha  \right),\\
\Gamma[\tilde \Phi,\Phi,\tilde N_5,N_5]&\equiv W[\tilde \Phi,\Phi,\tilde \mu_5,\mu_5]+ \int \dd t \int \dd \rv \left(\tilde \mu_5 \tilde N_5+\mu_5 N_5 \right).
\end{align}
One can show the following identities which will be used later (see, e.g., section\,4.\,4 of ref.~\cite{Tauber} for the derivations):
\begin{align}
\label{eq:id1}
\frac{\delta W}{\delta \Phi_\alpha}&=
\frac{\delta \Gamma}{\delta \Phi_\alpha}=j_\alpha,\quad
\frac{\delta \Gamma}{\delta N_5}=\mu_5,\\
\label{eq:id2}
\frac{\delta^2 \Gamma}{\delta \tilde \Phi_\alpha (\kv,\omega) \delta \Phi_\beta (\kv,\omega)}&=G^{-1}_{\alpha\beta}(-\kv,-\omega),\\
\label{eq:id3}
\frac{\delta^2 \Gamma}{\delta \Phi_\alpha  \delta  \Phi_\beta }&=\frac{\delta^2 \Gamma}{\delta \Phi_\alpha \delta N_5}=0,\\
\label{eq:id4}
\frac{\delta^3 \Gamma}{\delta N_5\delta \tilde \Phi_\alpha \delta \Phi_\beta}&=V_{\alpha;\beta 2}.
\end{align}
Here, note that $V_{\alpha;\beta 2}$ in eq.~(\ref{eq:id4}) satisfies eq.~(\ref{eq:mathcalVdef}).

We use the condition that $W[\tilde \Phi,\Phi,\tilde \mu_5,\mu_5]$ is invariant under the $\text{U(1)}_\text A^3$ transformation. Note that $W[\tilde \Phi,\Phi,\tilde \mu_5,\mu_5]$ is a functional of $\mu_5$, and changing $\mu_5$ corresponds to the $\text{U(1)}_\text A^3$ transformation, because $n_5$ is its generator. Suppose we turn on a variation of $\mu_5$ at $t=0$, $\delta \mu_5=\vartheta\mu_5$ with $\vartheta$ being a small parameter, so that the following contribution is added to the free energy~(\ref{eq:GL}):
\begin{align}
\delta F=\int \dd\rv\, n_5 \delta\mu_5.
\end{align}
Then, by using eq.~(\ref{eq:Langephi}), an infinitesimal $\text{U(1)}_\text A^3$ transformation can be written as
\begin{align}
\label{eq:U(1)Amu5} 
\delta  \Phi_\alpha = g\int_0^t \dd t' \varepsilon_{\alpha\beta} \Phi_\beta \delta \mu_5 = \vartheta g\varepsilon_{\alpha\beta}\Phi_\beta \mu_5 t.
\end{align}
By applying eq.~(\ref{eq:U(1)Amu5}) to $W[\tilde \Phi,\Phi,\tilde \mu_5,\mu_5]$, we obtain its variation as
\begin{align}
\delta W&= \vartheta \int \dd t \int \dd \rv \,  \mu_5  \left[ \frac{\delta W}{\delta \mu_5} + g  \varepsilon_{\alpha\beta} t \frac{\delta W}{\delta \Phi_\alpha}\Phi_\beta  \right]\\
&= \vartheta \int \dd t \int \dd \rv \, \frac{\delta \Gamma}{\delta N_5}  \left[  N_5 + g  \varepsilon_{\alpha\beta} t \frac{\delta \Gamma}{\delta \Phi_\alpha}\Phi_\beta   \right],
\end{align}
where we have used eq.~(\ref{eq:id1}). 
Therefore, invariance of $W[\tilde \Phi, \Phi,\tilde \mu_5,\mu_5]$
leads to 
\begin{align}
\frac{\delta \Gamma}{\delta N_5}  \left[  N_5 + g  \varepsilon_{\alpha\beta} t \frac{\delta \Gamma}{\delta \Phi_\alpha}\Phi_\beta  \right]=0.
\end{align}
By taking a variation of this equation with respect to $\tilde \Phi_\alpha$ and $\Phi_\beta$, we obtain
\begin{align}
\label{eq:WT}
\frac{\delta^3 \Gamma}{\delta N_5\delta \tilde \Phi_\alpha \delta \Phi_\beta}N_5=-g\varepsilon_{\gamma\beta} t \left( \mu_5 \frac{ \delta^2 \Gamma}{\delta \tilde \Phi_\alpha  \delta \Phi_\gamma}+ j_\gamma \frac{\delta^2 \Gamma}{\delta N_5 \delta\tilde \Phi_\alpha } \right),
\end{align}
where we have used eqs.~(\ref{eq:id1}) and (\ref{eq:id3}). 
Then, we take the functional derivative of eq.~(\ref{eq:WT}) with respect to $\mu_5$ and set all the external fields to zero, yielding the following result in the frequency space,
\begin{align}
\frac{\delta^3 \Gamma}{\delta N_5\delta \tilde \Phi_\alpha \delta \Phi_\beta}=-i\frac{g\varepsilon_{\gamma\beta}}{\chi_5} \left.\frac{\partial }{\partial \omega}\frac{ \delta^2 \Gamma}{\delta \tilde \Phi_\alpha  \delta \Phi_\gamma}\right|_{\omega\rightarrow0}
=\frac{g\varepsilon_{\gamma\beta}}{\chi_5} \left( 1-i \left.\frac{\partial \Sigma_{\alpha\gamma}(\zev,\omega)}{\partial \omega}\right|_{\omega\rightarrow0}\right).
\end{align}
Here, we have used $\chi_5=\delta N_5/\delta \mu_5$ and eqs.~(\ref{eq:id2}) and (\ref{eq:DysonG}). By using eqs.~(\ref{eq:id4}), (\ref{eq:bareV}) and (\ref{eq:mathcalVdef}), we finally arrive at eq.~(\ref{eq:V}) when $\alpha\neq\beta$. Note here that one can write $\varepsilon_{\gamma\beta}\Sigma_{\alpha\gamma}=\varepsilon_{\alpha\beta}\Sigma_{11}\ ({\rm for}\ \alpha\neq\beta)$ by using $\Sigma_{11}=\Sigma_{22}$, which can be shown from eq.~(\ref{eq:Sigma_alal}).

\section{Crossover phenomena between different dynamic universality classes}
\label{sec:crossover}
In this appendix, we discuss the crossover phenomena of the critical behaviors as functions of the relevant parameters between the fixed points (iii) and (iv) (or (i) and (iv)) considered in section \ref{sec:stability}.

We first show that the parameter region that leads to the fixed point (iv) under RG is much larger than (iii) within the linear-stability analysis around the fixed point (i). 
Let us look at the full propagator of the conserved fields, $D_{ij}=\bar D_{ij}(w,h)$ as a function of two relevant parameters $w$ and $h$, which is transformed under the rescaling as 
\begin{align}
\label{eq:D1}
\bar D_{ij}(w,h)\sim \bar D_{ij}(l^{y_w}w,l^{y_h}h).
\end{align}
Here, we assume that $w$ and $h$ are scaled by the eigenvalues $y_w\equiv 7\epsilon/10,\,y_h\equiv 1-\epsilon/4$ obtained in eq.~(\ref{eq:eigenvalues}), and we omit the overall factor. When the RG flow approaches $l\sim h^{-1/y_h}$, $\bar D_{ij}$ becomes a function of $wh^{-\varphi}$ with $\varphi\equiv y_w/y_h$ being the crossover exponent, and it behaves differently depending on whether $wh^{-\varphi}\gg 1$ or $wh^{-\varphi}\ll 1$: in the former region, we see the behavior of the fixed point (iii); in the latter region, we see the behavior of the fixed point (iv). Since $\varphi=O(\epsilon)$ and $w\ll1$ near the fixed point (i), the latter parameter region is much larger than the former. These two regions should be continuously connected, because the right-hand side of the original RG equations (\ref{eq:REf})--(\ref{eq:REh}) are smooth functions of all the parameters, unless $w$ and $h$ are simultaneously infinity.

In addition, by considering the initial parameter region with $w \ll 1$, we also find the crossover between the fixed points (i) and (iv) depending on the magnitude of the magnetic field $B$ and the reduced temperature $\tau$ as follows. We recall that $r\approx \tau$ is also a relevant parameter. Similarly to the discussion in the previous paragraph, RG transformation from the fixed point (i) by $l\sim \tau^{-1/\nu}$ yields the propagator as a function of $\tau h ^{-\varphi'}$. Here, $\nu$ defined by eq.~(\ref{eq:eps}) is the critical exponent of $\tau$, and $\varphi'\equiv(\nu y_h)^{-1}$ is the crossover exponent. Therefore, the dynamic universality class belongs to that of model A governed by the fixed point (iv)
in the regime $\tau\ll h^{\varphi'}$ and to that of model E governed by the fixed point (i) in the regime $\tau\gg h^{\varphi'}$ for a fixed magnetic field.


\begin{thebibliography}{99}
  \bibitem{Adler}
  S.~Adler,
  Phys.\ Rev.\ {\bf 177}, 2438 (1969).

  \bibitem{BellJackiw}
  J.~S.~Bell and R.~Jackiw,
  Nuovo Cimento {\bf 60A}, 47 (1969).

 \bibitem{Kharzeev:2007jp} 
  D.~E.~Kharzeev, L.~D.~McLerran and H.~J.~Warringa,
  Nucl.\ Phys.\ A {\bf 803}, 227 (2008).
  
  \bibitem{Fukushima:2008xe} 
  K.~Fukushima, D.~E.~Kharzeev and H.~J.~Warringa,
  Phys.\ Rev.\ D {\bf 78}, 074033 (2008).
  
  \bibitem{Nielsen:1983rb} 
  H.~B.~Nielsen and M.~Ninomiya,
  Phys.\ Lett.\  {\bf 130B}, 389 (1983).

 \bibitem{Vilenkin:1980fu} 
  A.~Vilenkin,
  Phys.\ Rev.\ D {\bf 22}, 3080 (1980).
  
  \bibitem{Vilenkin:1979ui} 
  A.~Vilenkin,
  Phys.\ Rev.\ D {\bf 20}, 1807 (1979).
  
  \bibitem{Kharzeev:2007tn} 
  D.~Kharzeev and A.~Zhitnitsky,
  Nucl.\ Phys.\ A {\bf 797}, 67 (2007).
    
  \bibitem{Son:2009tf}
  D.~T.~Son and P.~Sur\'owka,
  Phys.\ Rev.\ Lett.\  {\bf 103}, 191601 (2009).
  
  \bibitem{Landsteiner:2011cp} 
  K.~Landsteiner, E.~Megias and F.~Pena-Benitez,
  Phys.\ Rev.\ Lett.\  {\bf 107}, 021601 (2011).
  
  \bibitem{Kharzeev:2010gd} 
  D.~E.~Kharzeev and H.~U.~Yee,
  Phys.\ Rev.\ D {\bf 83}, 085007 (2011).
  
  \bibitem{Newman:2005hd} 
  G.~M.~Newman,
  JHEP {\bf 0601}, 158 (2006).

  \bibitem{Burnier:2011bf}
  Y.~Burnier, D.~E.~Kharzeev, J.~Liao and H.~U.~Yee,
  Phys.\ Rev.\ Lett.\  {\bf 107}, 052303 (2011).
  
  \bibitem{Adamczyk:2015eqo}
  L.~Adamczyk {\it et al.} [STAR Collaboration],
  Phys.\ Rev.\ Lett.\  {\bf 114}, 252302 (2015).
  
  \bibitem{Hatta:2015hca} 
  Y.~Hatta, A.~Monnai and B.~W.~Xiao,
  Nucl.\ Phys.\ A {\bf 947}, 155 (2016).

  \bibitem{Hongo:2013cqa} 
  M.~Hongo, Y.~Hirono and T.~Hirano,
  Phys.\ Lett.\ B {\bf 775}, 266 (2017).
  
  \bibitem{Stephanov:2004wx} 
  M.~A.~Stephanov,
  Prog.\ Theor.\ Phys.\ Suppl.\  {\bf 153}, 139 (2004)
  [Int.\ J.\ Mod.\ Phys.\ A {\bf 20}, 4387 (2005)].
  
  \bibitem{Asakawa:1989bq}
  M.~Asakawa and K.~Yazaki,
  Nucl.\ Phys.\ {A\bf 504}, 668 (1989).
  
  \bibitem{Barducci:1989wi}
  A.~Barducci, R.~Casalbuoni, S.~De Curtis, R.~Gatto and G.~Pettini,
  Phys.\ Lett.\ B {\bf 231}, 463 (1989).
  
  \bibitem{Kitazawa:2002bc}
  M.~Kitazawa, T.~Koide, T.~Kunihiro and Y.~Nemoto,
  Prog.\ Theor.\ Phys.\  {\bf 108}, 929 (2002).
  
  \bibitem{Hatsuda:2006ps}
  T.~Hatsuda, M.~Tachibana, N.~Yamamoto and G.~Baym,
  Phys.\ Rev.\ Lett.\  {\bf 97}, 122001 (2006); 
  Phys.\ Rev.\ D {\bf 76}, 074001 (2007).
  
  \bibitem{Zhang:2008wx} 
  Z.~Zhang, K.~Fukushima and T.~Kunihiro,
  Phys.\ Rev.\ D {\bf 79}, 014004 (2009).

  \bibitem{Rajagopal:1992qz} 
  K.~Rajagopal and F.~Wilczek,
  Nucl.\ Phys.\ B {\bf 399}, 395 (1993).
  
  \bibitem{Fujii:2003bz} 
  H.~Fujii,
  Phys.\ Rev.\ D {\bf 67}, 094018 (2003).
  
  \bibitem{Fujii:2004jt} 
  H.~Fujii and M.~Ohtani,
  Phys.\ Rev.\ D {\bf 70}, 014016 (2004).
  
  \bibitem{Son:2004iv} 
  D.~T.~Son and M.~A.~Stephanov,
  Phys.\ Rev.\ D {\bf 70}, 056001 (2004).
  
  \bibitem{Minami:2011un} 
  Y.~Minami,
  Phys.\ Rev.\ D {\bf 83}, 094019 (2011).
  
  \bibitem{Hohenberg:1977ym} 
  P.~C.~Hohenberg and B.~I.~Halperin,
  Rev.\ Mod.\ Phys.\  {\bf 49}, 435 (1977).
  
  \bibitem{Sogabe:2016ywr} 
  N.~Sogabe and N.~Yamamoto,
  Phys.\ Rev.\ D {\bf 95}, 034028 (2017).

  \bibitem{Jiang:2015cva} 
  Y.~Jiang, X.~G.~Huang and J.~Liao,
  Phys.\ Rev.\ D {\bf 92}, 071501 (2015).
  
  \bibitem{Onuki:1997} 
  A.~Onuki,
  Phys.\ Rev.\ E {\bf 55}, 403 (1997).
  
  \bibitem{Pankert:1986}  
  J.~Pankert and V.~Dohm,
  Europhys.\ Lett.\ {\bf 2}, 775 (1986); 
  Phys.\ Rev.\ B\ {\bf 40}, 10842 (1989); Phys.\ Rev.\ B\ {\bf 40},10856 (1989).
  
  \bibitem{Martin:1973zz} 
  P.~C.~Martin, E.~D.~Siggia and H.~A.~Rose,
  Phys.\ Rev.\ A {\bf 8}, 423 (1973).

  \bibitem{Janssen:1976} 
  H-K. Janssen,
  Z.\ Phys.\ B {\bf 23}, 377 (1976).

  \bibitem{DeDominicis:1978} 
  C. De Dominicis,
  Phys.\ Rev.\ B {\bf 18}, 4913 (1978).

  \bibitem{Shushpanov:1997sf} 
  I.~A.~Shushpanov and A.~V.~Smilga,
  Phys.\ Lett.\ B {\bf 402}, 351 (1997).

  \bibitem{Pisarski:1983ms} 
  R.~D.~Pisarski and F.~Wilczek,
  Phys.\ Rev.\ D {\bf 29}, 338 (1984).
  
  \bibitem{Chaikin}
  P.~M.~Chaikin and T.~C.~Lubensky,
  {\it Principles of Condensed Matter Physics}
  (Cambridge University Press, Cambridge, England, 1995).

  \bibitem{Forster}
  D.~Forster,
  {\it Hydrodynamic fluctuations, broken symmetry, and correlation functions}
  (Perseus Books, New York, 1975).
  
  \bibitem{Jackiw}
  R.~Jackiw, ``Field theoretic investigations in current algebra,'' in
  {\em Lectures on Current Algebra and Its Applications,}
  ed.\ S.~B.~Treiman, R.~Jackiw and D.~J.~Gross (Princeton University
  Press, Princeton, NJ, 1972).

\bibitem{Faddeev:1984jp}
  L.~D.~Faddeev,
  Phys.\ Lett.\ B {\bf 145}, 81 (1984).
  
  \bibitem{Son:2012wh} 
  D.~T.~Son and N.~Yamamoto,
  Phys.\ Rev.\ Lett.\  {\bf 109}, 181602 (2012).

  \bibitem{Tauber}  
  U.~C.~T\"auber, {\it Critical dynamics: a field theory approach to equilibrium and
  non-equilibrium scaling behavior} (Cambridge University Press, Cambridge, 2014).

  \bibitem{Halperin:1974zz} 
  B.~I.~Halperin, P.~C.~Hohenberg and S.~k.~Ma,
  Phys.\ Rev.\ B {\bf 10}, 139 (1974).

  \bibitem{Halperin:1976zza} 
  B.~I.~Halperin, P.~C.~Hohenberg and E.~D.~Siggia, 
  Phys.\ Rev.\ B {\bf 13}, 1299 (1976).

  \bibitem{Onuki:PTD}  
  A.~Onuki, {\it Phase Transition Dynamics}, (Cambridge University Press, 2007).
  
  \bibitem{Golkar:2012kb} 
  S.~Golkar and D.~T.~Son,
  JHEP {\bf 1502}, 169 (2015).

   \bibitem{Akamatsu:2013pjd} 
  Y.~Akamatsu and N.~Yamamoto,
  Phys.\ Rev.\ Lett.\  {\bf 111}, 052002 (2013).

 \bibitem{ONWT}
 U.~C.~T\"auber and Z.~R\'{a}cz, 
 Phys.\ Rev.\ E {\bf 55}, 4120 (1997).
 
\end{thebibliography}
\end{document}